\documentclass[sigconf]{acmart}
\usepackage{natbib}
\usepackage{mathtools}
\usepackage{multicol, multirow}
\usepackage{threeparttable}
\usepackage{booktabs}
\usepackage{amsmath}

\newtheorem{myDef}{Req.}

\usepackage{subfigure}
\usepackage{algorithm}
\usepackage{algpseudocode}
\usepackage{ulem}
\usepackage{caption}
\usepackage{bbding}
\usepackage{enumitem}



\copyrightyear{2022}
\acmYear{2022}
\setcopyright{acmcopyright}\acmConference[CIKM '22]{Proceedings of the 31st ACM International Conference on Information and Knowledge Management}{October 17--21, 2022}{Atlanta, GA, USA}
\acmBooktitle{Proceedings of the 31st ACM International Conference on Information and Knowledge Management (CIKM '22), October 17--21, 2022, Atlanta, GA, USA}
\acmPrice{15.00}
\acmDOI{10.1145/3511808.3557411}
\acmISBN{978-1-4503-9236-5/22/10}

\begin{document}

\title{OptEmbed: Learning Optimal Embedding Table for Click-through Rate Prediction}

\author{Fuyuan Lyu}
\authornote{Both authors contributed equally to this research.}
\affiliation{
  \institution{McGill University}
  \city{Montreal}
  \country{Canada}
}
\email{fuyuan.lyu@mail.mcgill.ca}

\author{Xing Tang}
\authornotemark[1]
\authornote{Corresponding authors}
\email{xing.tang@huawei.com}
\affiliation{
  \institution{Huawei Noah's Ark Lab}
  \city{Shenzhen}
  \country{China}
}

\author{Hong Zhu}
\email{zhuhong8@huawei.com}
\author{Huifeng Guo}
\email{huifeng.guo@huawei.com}
\affiliation{
  \institution{Huawei Noah's Ark Lab}
  \city{Shenzhen}
  \country{China}
}

\author{Yingxue Zhang}
\affiliation{
  \institution{Huawei Noah's Ark Lab}
  \city{Montreal}
  \country{Canada}
}
\email{yingxue.zhang@huawei.com}

\author{Ruiming Tang}
\authornotemark[2]
\affiliation{
  \institution{Huawei Noah's Ark Lab}
  \city{Shenzhen}
  \country{China}
}
\email{tangruiming@huawei.com}

\author{Xue Liu}
\affiliation{
  \institution{McGill University}
  \city{Montreal}
  \country{Canada}
}
\email{xueliu@cs.mcgill.ca}

\renewcommand{\shortauthors}{Fuyuan Lyu et al.}

\begin{abstract}
Click-through rate (CTR) prediction model usually consists of three components: embedding table, feature interaction layer, and classifier. Learning embedding table plays a fundamental role in CTR prediction from the view of the model performance and memory usage. The embedding table is a two-dimensional tensor, with its axes indicating the number of feature values and the embedding dimension, respectively. To learn an efficient and effective embedding table, recent works either assign various embedding dimensions for feature fields and reduce the number of embeddings respectively or mask the embedding table parameters. However, all these existing works cannot get an optimal embedding table. On the one hand, various embedding dimensions still require a large amount of memory due to the vast number of features in the dataset. On the other hand, decreasing the number of embeddings usually suffers from performance degradation, which is intolerable in CTR prediction. Finally, pruning embedding parameters will lead to a sparse embedding table, which is hard to be deployed. To this end, we propose an optimal embedding table learning framework OptEmbed, which provides a practical and general method to find an optimal embedding table for various base CTR models. Specifically, we propose pruning the redundant embeddings regarding corresponding features' importance by learnable pruning thresholds. Furthermore, we consider assigning various embedding dimensions as one single candidate architecture. To efficiently search the optimal embedding dimensions, we design a uniform embedding dimension sampling scheme to equally train all candidate architectures, meaning architecture-related parameters and learnable thresholds are trained simultaneously in one supernet. We then propose an evolution search method based on the supernet to find the optimal embedding dimensions for each field. Experiments on public datasets show that OptEmbed can learn a compact embedding table which can further improve the model performance.
\end{abstract}

\maketitle
\section{Introduction}

\begin{figure}[!htbp]
    \centering
    \centering
    \includegraphics[width=0.4\textwidth]{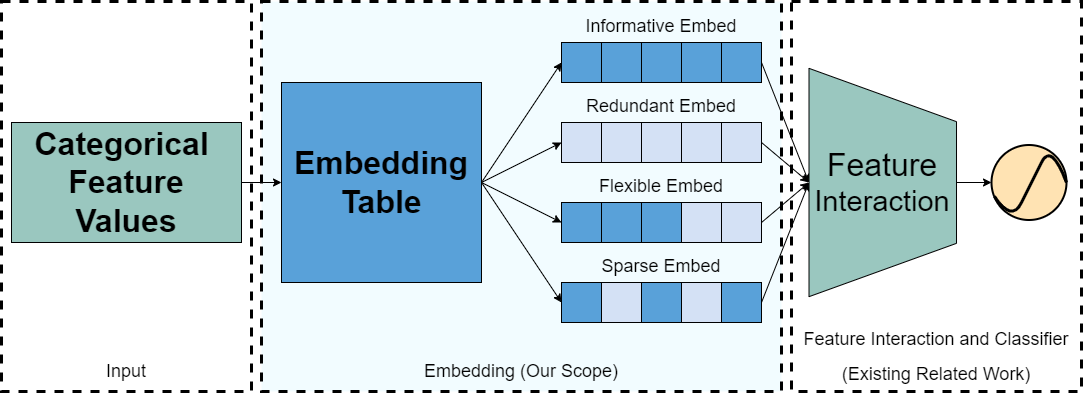}
    \vspace{-5pt}
    \caption{Overview of general framework of CTR prediction.}
    \label{fig:normal}
    \vspace{-10pt}
\end{figure}

Click-through rate (CTR) prediction has been a critical task in real-world commercial recommender systems and online advertising systems~\cite{ADS,DLRM}. It aims to predict the probability of a certain user clicking a recommended item (e.g. movie, advertisement)~\cite{FM,DeepFM,DCN,AutoInt}. General CTR prediction model architecture consists of embedding table, interaction layer, and classifier as illustrated in Fig. \ref{fig:normal}~\cite{AutoIAS,DeepFM,DCN,AutoInt}. The typical inputs of CTR models consist of many categorical features. We term the values of these categorical features as feature values, which are organized as feature fields. For example, a feature field \textit{gender} contains three feature values, \textit{male}, \textit{female} and \textit{unknown}. These predictive models use the embedding table to map the categorical feature values into real-valued dense vectors. Then these embeddings are fed into the feature interaction layer, such as factorization machine~\cite{FM}, cross network~\cite{DCN}, self-attention layer~\cite{AutoInt}. The final classifier aggregates the representation vector to make the prediction. 

In the general CTR prediction model architecture, the embedding table dominates the number of parameters and plays a fundamental role in prediction performance. Therefore, it is critical to obtain optimal embedding tables that reduce the model size and improve performance~\cite{MGQE,QR}. The embedding table is a two-dimensional tensor $\mathbf{E} \in R^{|f|\times D}$, which maps each feature value to an indexed row. The first dimension size $|f|$ thus equals the total number of feature values, and the second one $D$ is the embedding dimension. The memory cost of the embedding table is $O(|f|D)$, where $f$ mainly decides the memory usage given $|f| \gg D$. When all possible feature values are fed into the model~\cite{DeepFM,NFM}, $|f|$ becomes a vast number (up to millions on web-scale applications). This leads to a large number of embeddings, which contributes to the primary memory bottleneck within both training and inference ~\cite{sfctr}. However, the redundant embedding not only necessitates additional memory cost but is also detrimental to the model performance~\cite{autofield}. Therefore, the first requirement for an optimal embedding table is to distinguish the redundant embeddings and zero them out before they are fed into the following layers, as shown in Fig. \ref{fig:normal}.
The embedding dimension $D$ is mostly fixed across all the feature values. Previous works\cite{MGQE,AutoDim} point out that the over-parameterizing features with smaller feature cardinality may induce overfitting, and features with larger cardinality need larger dimensions to convey fruitful information. Hence, the second requirement for an optimal embedding table is to assign various embedding dimensions to feature values as flexible embedding shown in Fig. \ref{fig:normal}.
An alternative to optimize the embedding table is to mask the embedding parameters directly \cite{single-shot, PEP}. It makes embedding dimension $D \geq 0$ with discontinuous parameters, illustrated as sparse embedding in Fig. \ref{fig:normal}. Nevertheless, the sparse embedding table requires storing extra structural information and additional computation cost in the inference stage, which is not suitable for hardware in practical ~\cite{deeplight}.
The final requirement is to optimize the embedding table without storing additional structural information on hardware requirements. 
Given the above requirements, it is highly desired to prune redundant embeddings and search embedding dimensions in a unifying way.

Previous work on optimizing the embedding table either treats embedding reduction and dimensions search separately or generates sparse embedding. To reduce the number of embeddings, one natural approach is to design a hash function, which maps the categorical features to the embedding table index ~\cite{featurehashing, doublehashing, binarycode}. Since the embedding table index size is far less than the number of feature values, this approach optimizes memory usage. However, blindly mapping different feature values into the same embeddings without distinguishing the redundant embeddings may lead to performance degradation, which is intolerable in CTR prediction. On the other hand, AutoField~\cite{autofield} utilizes the differential architecture search~\cite{DARTS} method to prune redundant feature fields. But this method may prune some informative feature values while preserving some redundant ones since the feature field is not fine-grained enough to generate an optimal embedding. 
To search for flexible embedding dimension, AutoDim~\cite{AutoDim} also utilize the differential architecture search~\cite{DARTS} method. However, this method cannot get an optimal embedding table, as it does not remove redundant features. 
Moreover, some research optimizes the embedding table in a unifying way based on embedding pruning~\cite{PEP,UMEC,single-shot}. These methods identify and mask redundant values in embeddings, where embeddings are pruned when their dimensions are equal to zero. However, these methods result in a sparse embedding table, which poses challenges when fitting into modern computation units.

In this paper, we propose a framework to address two main challenges and learn an \textbf{opt}imal \textbf{embed}ding table (OptEmbed) that satisfy all three requirements. First, for the problem of \textbf{how to prune the redundant embeddings and search feature fields embedding dimensions in a unifying way}, we transform it into the problem of identifying the importance of each feature value. Besides, searching field-wise embedding dimensions can be formulated as an architecture search problem. To this end, inspired by structural pruning and network architecture search~\cite{LTH,proxylessnas}, we introduce learnable pruning thresholds to distinguish informative embeddings and to allocate the dimensions to those embeddings in an automated and data-driven manner. Specifically, we mask redundant embeddings adaptively with thresholds. Meanwhile, we design a uniform embedding dimension sampling scheme to train a supernet with the learnable thresholds. Then, We conduct an evolutionary search based on the supernet with informative embeddings to assign optimal field-wise dimensions. To address the second challenge of \textbf{how to optimize the embedding table efficiently}, we reparameterize the problem with a threshold vector, which makes the original problem differentiable and only needs a few preallocate memory~\cite{featureselection}. Moreover, the search space of embedding dimensions is also too huge to explore~\cite{AutoDim}. Therefore, we design a one-shot embedding dimension search method to save search time by decoupling parameter training and dimension search based on the supernet mentioned above. The experimental results on three public datasets demonstrate the efficiency and effectiveness of our proposed framework OptEmbed. We summarize our major contributions as below:
\begin{itemize}[topsep=0pt,noitemsep,nolistsep,leftmargin=*]
    \item This paper firstly proposes the requirements for an optimal embedding table: no redundant embedding, embedding dimension flexible and hardware friendly. We propose a novel optimization method called OptEmbed, which improves model performance and reduces memory usage based on these requirements.
    \item The proposed OptEmbed optimizes the embedding table in a unifying way. It can efficiently train a supernet with informative feature values and the embedding parameters simultaneously. Moreover, we design a one-shot embedding dimension search method based on the supernet, which produces the optimal embedding table without sparse embedding.
    \item The extensive experiments are conducted on three public datasets. The experimental results demonstrate the effectiveness and efficiency of the proposed framework.
\end{itemize} 

We organize the rest of this paper as follows. In Section \ref{sec:problem}, we formulate the CTR prediction problem and three requirements for the optimal embedding table. In Section \ref{sec:optembed}, we present OptEmbed to obtain the optimal embedding table efficiently. Section \ref{sec:experiment} details the experiments. In Section \ref{sec:rw}, we briefly introduce related works. Finally, we conclude this work in Section \ref{sec:conclusion}.
\section{Problem Definition}
\label{sec:problem}
In this section, we formulate how the CTR prediction model output the prediction result with the concatenation of multiple features and define the requirements for an optimal embedding table.

\subsection{CTR Prediction}
\label{sec:problem:ctr}


We represent the raw inputs as the raw feature vector that concatenates $n$ feature fields $\mathbf{x}=[\mathbf{x}_{(1)},  \mathbf{x}_{(2)}, \cdots, \mathbf{x}_{(n)}]$. Usually, $\mathbf{x}_{(i)}$ is a one-hot representation, which is very sparse and high-dimensional. For example, the feature field \textit{gender} has three unique feature values, \textit{male},\textit{female}, and \textit{unknown}, then they can be represented by three one-hot vectors $[1,0,0]$,$[0,1,0]$ and $[0,0,1]$, respectively. Before raw feature vectors are fed into the feature interaction layer, we usually employ embedding table to convert them into low dimensional and dense real-value vectors. This can be formulated as $\mathbf{e}_{(i)}=\mathbf{E} \times \mathbf{x}_{(i)}, 1 \le i \le n$, where $\mathbf{E}\in\mathbb{R}^{|f|\times D}$ is the embedding table, $|f|$ is the number of feature values and $D$ is the size of embedding. Then embeddings are stacked together as a embedding vector $\mathbf{
e} = [\mathbf{e}_{(1)}, \mathbf{e}_{(2)}, \cdots, \mathbf{e}_{(n)}]$.

Following learnable embedding table, the feature interaction layer will be performed based on $\mathbf{e}$ in mainstream CTR models. There are several types of feature interaction in previous study, e.g. inner product~\cite{DeepFM}. As discussed in previous work~\cite{AutoPI}, feature interaction can be defined as based on embeddings:
\begin{equation}
    \mathbf{v}^p = o^{(p-1)}(o^{(p-2)}(\cdots(o^{(1)}(\mathbf{e}))\cdots)),
    \label{eq:interaction}
\end{equation}
where $o$ can be a single layer perceptron or cross layer\cite{DCN}. The feature interaction can be aggregated together:
\begin{equation}
    \hat{y} = \sigma(\mathbf{w}^T(\mathbf{v}^{(1)}\oplus\mathbf{v}^{(2)}\oplus\cdots\oplus\mathbf{v}^{(n)})+b) = \mathcal{F}(\mathbf{E} \times \mathbf{x}|\mathbf{W}),
    \label{eq:aggregation}
\end{equation}
where symbol $\oplus$ denotes the concatenation operation, $\mathbf{v}^{(k)}$ is the output of feature interaction, and $\mathbf{W}$ is network parameters except for embedding table. The cross entropy loss (i.e. log-loss) is adopted for training the model:
\begin{equation}
\label{eq:logloss}
    \text{CE} (y,\hat{y}) = y\log(\hat{y}) + (1-y)\log(1-\hat{y}).
\end{equation}
We summarize the final CTR prediction problem as follows:
\begin{equation}
\label{eq:summarize}
    \min_{\mathbf{E}, \mathbf{W}} \ \mathcal{L}_{\text{CE}}(\mathcal{D}|\{\mathbf{E}, \mathbf{W}\}) = -\frac{1}{ |\mathcal{D}|} \sum_{(\mathbf{x}, y) \in \mathcal{D}} \text{CE}(y, \mathcal{F}(\mathbf{E} \times \mathbf{x}|\mathbf{W})).
\end{equation}
where $y$ is the ground truth of user clicks, $\mathcal{D}$ is the training dataset. 


\subsection{Optimal Embedding Table}
\label{sec:problem:optimal}

The original embedding table $\mathbf{E} \in \mathbb{R}^{|f| \times D}$ is neither effective nor efficient \cite{AutoDim,autofield,AutoIAS,PEP}. 
An optimal embedding table that satisfies the following requirements can greatly reduce the model size and improve performance:

\begin{myDef}
\label{def:r1}
\emph{No Redundant Embeddings}: The optimal embedding table should only map informative feature values to embeddings. Feature value $\mathbf{x}_i$ is considered informative if the performance of the model is degraded when masking its corresponding embedding $\mathbf{e}_i$. Otherwise, it is deemed to be redundant. 
\end{myDef}

\begin{myDef}
\label{def:r2} \emph{Embedding Dimension Flexible}: The optimal embedding table should assign various embedding dimensions, improving the performance of the predictive model the most.
\end{myDef}

\begin{myDef}
\label{def:r3} \emph{Hardware Friendly}: The optimal embedding table should be compatible with the modern parallel-processing hardware (e.g. GPU) -- requiring no additional resources when training and inference.

\end{myDef}

To fulfill the three requirements above, we decompose the original single embedding table $\mathbf{E}$ into a series of field-wise embedding tables $\mathbf{E}^{*} = [\mathbf{E}_{(1)}, \mathbf{E}_{(2)}, \cdots, \mathbf{E}_{(n)}]$, where $\mathbf{E}_{(i)} \in \mathbb{R}^{|f_{(i)}| \times D_{(i)}}$. To satisfy requirement (i), we prune some embeddings related to redundant feature values, which can be formulated as $\sum_{i=1}^n |f_{(i)}| \le |f|$. As to requirements (ii) and (iii), different embedding sizes for each field-wise embedding tables are allocated. In summary, an optimal embedding table can be further defined as:
\begin{equation}
\label{eq:summarize2}
\begin{aligned}
    & \min_{\mathbf{E}^*, \mathbf{W}} \mathcal{L}_{\text{CE}}(\mathcal{D}|\{\mathbf{E}^{*}, \mathbf{W}\}), \ \mathbf{E}^{*} = [\mathbf{E}_{(1)}, \mathbf{E}_{(2)}, \cdots, \mathbf{E}_{(n)}], \\
    s.t. & \ \mathbf{E}_{(i)} \in \mathbb{R}^{|f_{(i)}| \times D_{(i)}}, \ \sum_{i=1}^n |f_{(i)}| \le |f|, \ D_{(i)} \le D, \ \forall i \le n.
\end{aligned}
\end{equation}

Notes that previous methods can not satisfy all three requirements. We will detail this in Section \ref{sec:summary}.

\section{OptEmbed}
\label{sec:optembed}

\begin{figure}[!t]
    \centering
    \includegraphics[width=0.4\textwidth]{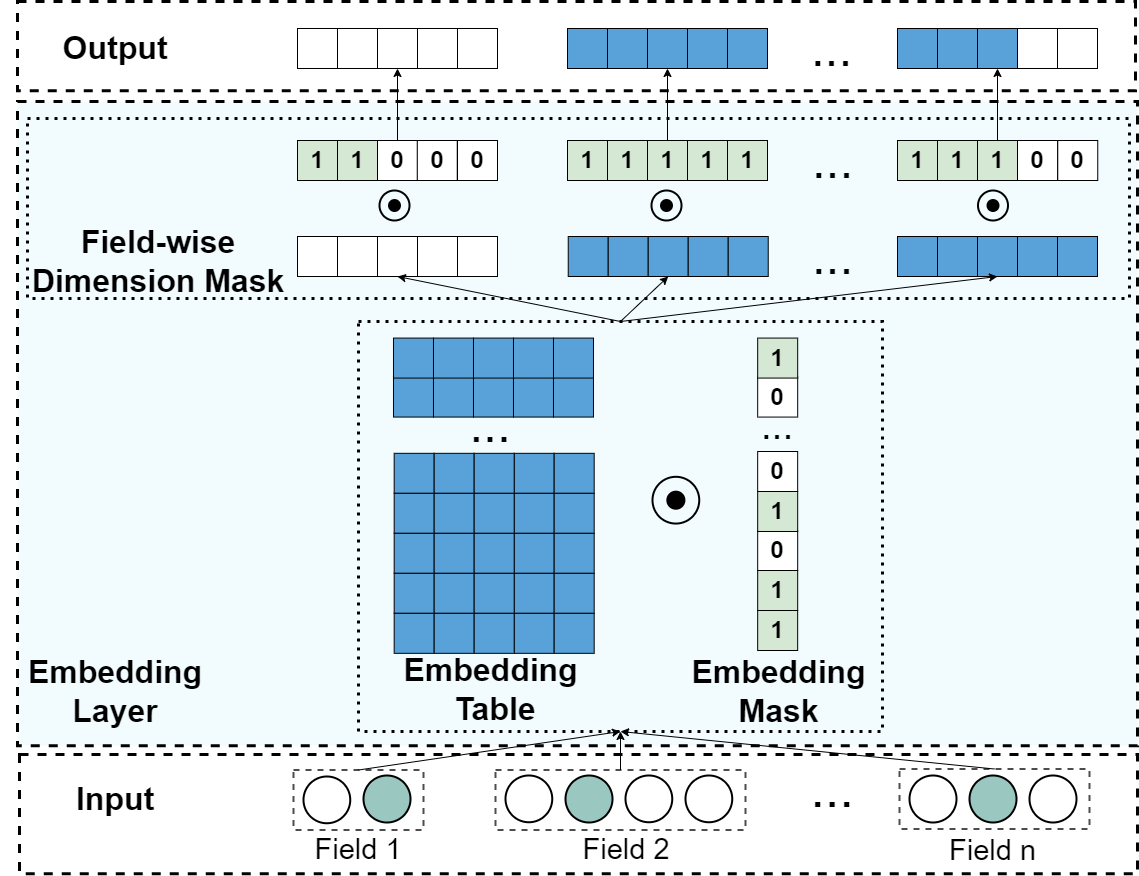}
    \vspace{-5pt}
    \caption{The Overview of OptEmbed.}
    \vspace{-10pt}
    \label{fig:overall}
\end{figure}

In this section, we propose a framework called OptEmbed to learn the optimal embedding table $\mathbf{E}^{*}$ defined in Section \ref{sec:problem}. We rewrite Eq. \ref{eq:summarize2} into the following by introducing two masks:

\begin{equation}
\label{eq:logloss2}
    \min_{\mathbf{m}_e, \mathbf{m}_d, \mathbf{E}, \mathbf{W}} \mathcal{L}_{\text{CE}}(\mathcal{D}|\{\mathbf{E}^{*}, \mathbf{W}\}) ,\ \mathbf{E}^{*} = \mathbf{E} \odot \mathbf{m}_e \odot \mathbf{m}_d.
\end{equation}
Here $\mathbf{m}_d \in \{0, 1\}^{D \times n}$ denotes the field-wise dimension mask. $\mathbf{m}_e \in \{0, 1\}^{|f|}$ denotes the embedding mask. $|f|$ and $n$ denote the feature number and field number. $\odot$ denotes element-wise product with broadcasting. By doing so, we decompose the task of learning the optimal embedding table $\mathbf{E}^{*}$ into three parts: (i) train embedding mask $\mathbf{m}^{*}_e$ to preserve informative embeddings; (ii) search for field-wise dimension mask $\mathbf{m}^{*}_d$ to assign various embedding size to each field and (iii) re-train the optimal embedding table $\mathbf{E}^{*}$ under the constraints of (i) and (ii). The overview of OptEmbed framework is shown in Fig. \ref{fig:overall}. Both the embedding mask and dimension mask are applied to the embedding table. Some of the embeddings are removed, while others become sparse.

Below, we first illustrate how to determine embedding mask $\mathbf{m}_e$ and field-wise dimension mask $\mathbf{m}_d$, respectively. We then introduce the re-training stage and discuss how OptEmbed compares with other methods that optimize the architecture of embedding tables from various aspects.

\subsection{Redundant Embedding Pruning}
\label{sec:feature_mask}

The redundant embedding pruning component determines which rows of the embedding table are informative and should be included in the final prediction task. Given that $|f|$ tends to be a large number, it is computationally inefficient to assign individual parameters to each feature marking its importance. Inspired by network pruning \cite{DST,Cont_Spar}, we directly optimize the embedding table $\mathbf{E}$ and adaptively pruning the embeddings via comparing with field-wise threshold, which can be updated by gradient descent. The reparameterization of the $\mathbf{m}_e$ is formulated as follows:
\begin{equation}
\label{eq:mask_f}
    \mathbf{m}_e = S(L_{\beta}(\mathbf{E})-\mathbf{t}),
\end{equation}
where $\mathbf{t}$ is the field-wise threshold vector, $L_{\beta}$ is the $\beta$ norm of embedding in each field, $S(\cdot)$ is the activation function, which works as trainable dynamic mask. We will illustrate three parts in the following sections in details.

\subsubsection{Field-wise Threshold Vector}
We introduce a trainable field-wise vector $\mathbf{t} \in \mathcal{R}^{|f|}$ to serve as pruning thresholds for embeddings in every field. We do not adopt a global threshold because corresponding features from different fields are likely to have different properties. For instance, the average frequency for the \textit{gender} field is expected to be much higher than that for the \textit{ID} field. Assessing the importance of features from different fields with a global threshold value would lead to a non-robust and hard-to-train network. Meanwhile, we do not adopt a feature-wise threshold vector considering it would significantly increase the number of total parameters and make it more likely to over-fit.

\begin{figure}[!htbp]
    \centering
    \includegraphics[width=0.4\textwidth]{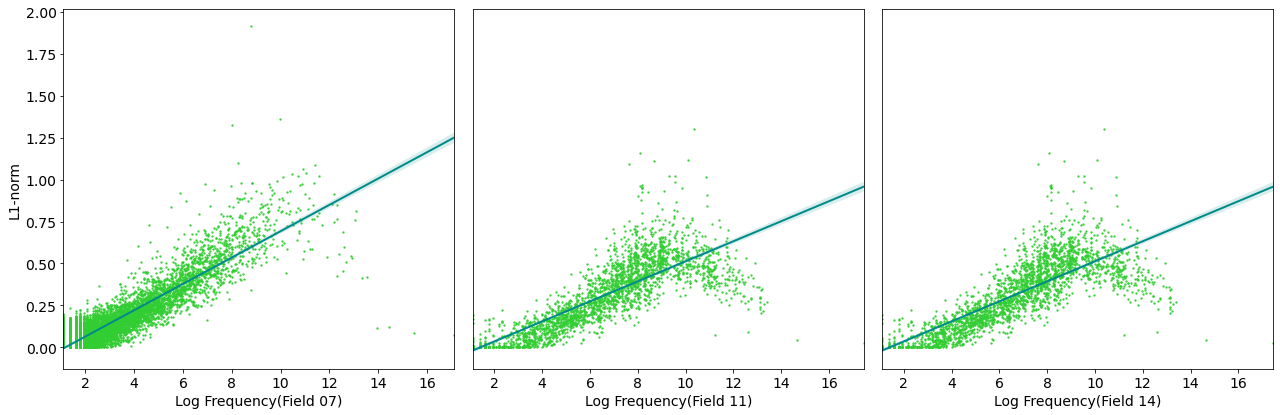}
    \vspace{-5pt}
    \caption{Relationship between $L_1$-norm and frequency in selected fields of Avazu dataset.}
    \vspace{-10pt}
    \label{fig:field}
\end{figure}

\subsubsection{$L_{\beta}$ norm}
It is commonly believed that features with higher frequency tend to be more important and informative in CTR prediction~\cite{AMTL,doublehashing}. To measure the importance of features precisely, we empirically train a prediction model and investigate the relation between the frequency of each feature and $L_{\beta}$ norm of the corresponding embedding in each field. Three fields in Avazu dataset\footnote{http://www.kaggle.com/c/avazu-ctr-prediction} are randomly selected as an example; here we set the base model as FNN~\cite{FNN}, $L_{\beta}$ norm as $L_1$ norm. The results are illustrated in Fig. \ref{fig:field}. Each green dot represents one embedding in the embedding table, with its x-axis denoting the $L_1$ norm of the embedding and y-axis denoting the log frequency of the corresponding feature value. A fitting curve is also shown in blue to summarize the relationships between these two variables. As we can observe, with the increment of feature frequency, the $L_1$ norm of the corresponding embedding also grows linearly. As a result, we adopt $L_{\beta}$ norm of corresponding embeddings as a measurement in our framework~\cite{DST,weight_connect}. 


\subsubsection{Unit Step Function}


We introduce a unit step function $S(x)$ as the activation function to generate a binary mask. 
Given the field-wise threshold vector $\mathbf{t}$ and unit step function $S(x)$, we can formally generate the embedding mask $\mathbf{m}_e$. For feature $j$ corresponding embedding $\mathbf{e}^j$, its embedding mask $\mathbf{m}_e^j$ is given by:
\begin{equation}
    \mathbf{m}_e^j = S(L_{\beta}(\mathbf{e}^j)-\mathbf{t}^{k_j}) = \left\{ 
    \begin{aligned}
        1&, \ L_{\beta}(\mathbf{e}^j)-\mathbf{t}^{k_j} > 0 \\
        0&, \ \text{otherwise}
    \end{aligned}
    \right. ,
\end{equation}
where $L_{\beta}(\cdot)$ indicates the $L_{\beta}$ normalization function and $k_j$ maps feature $j$ to the corresponding field. With the unit step function $S(x)$, we can easily generate binary embedding mask $\mathbf{m}_e$. Then the embedding table can be formulated as
\begin{equation}
\label{eq:embedding}
    \mathbf{\hat{E}} = \mathbf{E} \odot \mathbf{m}_e = \mathbf{E} \odot S(L_{\beta}(\mathbf{E})-\mathbf{t}).
\end{equation}

The prediction score will be calculated with $\mathbf{\hat{E}}$. However, because the derivative of step unit function is an impulse function, Eq.\ref{eq:embedding} cannot be directly optimized. To preserve gradients and make the model trainable, we adopt the long-tail derivation estimator~\cite{DST} to replace the gradient $dS(x)/{dx}$ of the step unit function. The long-tail derivation estimator can be formulated as
\begin{equation}
    \frac{d}{dx}S(x) \approx H(x) = \left\{ 
    \begin{aligned}
        &2-4|x|, &|x|\leq 0.4 \\
        &0.4, &0.4<|x|\leq 1 \\
        &0, &\text{otherwise}
    \end{aligned}
    \right. .
\end{equation}

We adopt this derivative long-tail estimator to optimize the field-wise threshold vector $\mathbf{t}$, so that the gradient would be large when the $L_{\beta}$ norm and the threshold value are close to each other and would be 0 when the gap between $L_{\beta}$ norm and threshold value are large enough. We denote the gradient of actual embedding $\mathbf{\hat{E}}$ as $d\mathbf{\hat{E}}$. Notes that the gradient of embedding for updating $\mathbf{E}$ is 
\begin{equation}
    d\mathbf{E} = d\mathbf{\hat{E}} \odot \mathbf{m}_e + d\mathbf{\hat{E}} \odot \mathbf{E} \odot H(L_{\beta}(\mathbf{E}) - \mathbf{t}) \odot dL_{\beta}(\mathbf{E}).
\end{equation}
The gradient of embedding is composed of two parts. The first part $d\mathbf{\hat{E}} \odot \mathbf{m}_e$ is the performance gradient that improves the performance. The second part $d\mathbf{\hat{E}} \odot \mathbf{E} \odot H(L_{\beta}(\mathbf{E}) - \mathbf{t}) \odot dL_{\beta}(\mathbf{E})$ is the structure gradient that removes redundant embeddings. As the final embedding is jointly influenced by both the performance and structure, OptEmbed can recover some embeddings. Specifically, once an embedding is accidentally removed, the performance gradient becomes zero because the embedding is zeroed-out. However, the embedding still receives the structure gradient. So the pruned embedding may be recovered again if the gap between $L_{\beta}$ norm and threshold are not too large (i.e. smaller than one in this case).


It is worth mentioning that the $\beta$ in $L_{\beta}$ norm needs to be carefully selected. With $L_{\beta}(\mathbf{e}) = (\sum e_i^{\beta})^{1/\beta}$ for embedding $\mathbf{e}$, we can get derivation for particular element:
\begin{equation}
\frac{dL_{\beta}}{de_i} = (\sum e_i^{\beta})^{1/\beta-1} \cdot e_i^{\beta-1}.
\end{equation}
Due to the first term $(\sum e_i^{\beta})^{1/\beta-1}$, the gradient of $e_i$ will be influenced by all elements from embedding $\mathbf{e}$ unless $\beta=1$. Therefore we select $\beta=1$ for OptEmbed hereafter to get rid of the influence.

\subsubsection{Sparse Regularization Term}

To remove more redundant embeddings, higher thresholds are encouraged. To achieve this, we explicitly add an exponential regularization term $L_s$ to the logloss that penalizes low threshold values. For the field-wise threshold $\mathbf{t} \in \mathbb{R}^{n}$, the exponential regularization term is 
\begin{equation}
    \mathcal{L}_s = \sum_{i=1}^n \exp(-t_i).
\end{equation}

Notice that the regularization term gradually decreases to zero as $x$ increases. Hence, the final objective in this stage becomes 
\begin{equation}
\label{eq:loss_feature}
    \min_{\mathbf{m}_e, \mathbf{E}, \mathbf{W}} \mathcal{L}_{\text{CE}}(\mathcal{D}|\{\hat{\mathbf{E}}, \mathbf{W}\}) + \alpha \mathcal{L}_s, \ \hat{\mathbf{E}} = \mathbf{E} \odot \mathbf{m}_e.
\end{equation}

Here $\alpha$ is the scaling coefficient for the sparse regularization term, controlling how many embeddings are pruned. With higher $\alpha$, $\mathcal{L}_s$ tends to increase the threshold $\mathbf{t}$, which makes it easier to prune redundant embeddings. However, once $\alpha$ becomes too large, it may accidentally remove certain informative embeddings, leading to the increase of the cross-entropy loss $\mathcal{L}_{\text{CE}}$. Therefore, our method can dynamically remove redundant embeddings, leading to a proper balance between model performance and size. 

\subsection{Embedding Dimension Search}
\label{sec:embed_mask}

The embedding dimension search component aims to assign various optimal dimensions for all fields. By viewing a group of field-wise dimension masks as one neural network architecture, we design an efficient neural architecture search method to search for optimal dimension masks efficiently in this section.

\subsubsection{One-shot NAS Problem}

Because the optimal embedding table should satisfy Req. \ref{def:r2} and \ref{def:r3}, the dimensionality set in our method, formed by all candidate embedding dimensions, can be formulated as $\mathcal{S}_e = \{1, \cdots, D-1, D\}$. Notice that the complexity of this search space is $O(D^n)$, which is impossible to search all the possible architectures in the search space exhaustively. On the other hand, to evaluate architecture, we need to train $\mathbf{m}_e$ and network parameters again, which costs a lot of computation resources. To efficiently search for the optimal embedding dimension mask, we hence re-formulate the dimension search as a one-shot NAS problem~\cite{One-shot,one-shotnas}:
\begin{equation}
\label{eq:one-shot}
\begin{aligned}
    \mathbf{m}_d^* & = \mathop{\arg\min}_{\mathbf{m}_d \in \mathcal{S}_e} \mathcal{L}_{\text{CE}}({\mathcal{D}_{val}|\{\hat{\mathbf{E}_s} \odot \mathbf{m}_d, \hat{\mathbf{W}_s}\}}), \\
    s.t. \ \{\hat{\mathbf{E}_s}, \hat{\mathbf{W}_s}\} & = \mathop{\arg\min}_{\{\mathbf{E}_s, \mathbf{W}_s\} \in \Omega} \mathbb{E}_{\mathbf{m}_d \sim \Gamma(\mathcal{S}_e)} \mathcal{L}_{\text{CE}}(\mathcal{D} | \{\mathbf{E}_s \odot \mathbf{m}_d , \mathbf{W}_s\}),
\end{aligned}
\end{equation}
where $\mathcal{S}_e$ denotes the search space, $\Gamma(\mathcal{S}_e)$ is the prior distribution of the search space, $\{\hat{\mathbf{E}_s}, \hat{\mathbf{W}_s}\}$ is the best supernet parameter and $\Omega$ denotes the parameter space of the supernet. By decoupling the dependency between training embedding and dimension search, we no longer need to train a sub-architecture from scratch, which reduces computation cost significantly. 

\subsubsection{Supernet Training} Following Eq. \ref{eq:one-shot}, we construct the supernet embedding table $\mathbf{E}_s$ with ordinal parameter sharing \cite{AutoIAS}, which is efficient to reuse most of parameters. In our method with search space $\mathcal{S}_e = \{1, \cdots, D-1, D\}$, the supernet is constructed with maximum dimension $D$. With no prior knowledge of $\Gamma$, we then assume $\Gamma$ as a uniform distribution. Such an assumption proves to be empirically good enough and efficient to apply~\cite{One-shot}. Specially, given $d \sim \text{Uniform}(1,D)$, the first $d$ elements of $\mathbf{m}_d$ are ones and the rest are zeros. Different from other methods~\cite{PEP,single-shot}, $\mathbf{m}_d$ induce flexible embedding, which is hardware-friendly, instead of sparse embedding, which requires a lot of structure information.

Moreover, the $\hat{\mathbf{E}}$ retains various embeddings during training both $\mathbf{m}_e$ and embedding parameters, which affects the supernet directly. To train the supernet adapting to $\mathbf{m}_e$ and reduce the total training time further, we conduct the supernet training and redundant embedding pruning in a unifying way by introducing $\mathbf{E}_s = \mathbf{E} \odot \mathbf{m}_e$. Finally, we can formulate supernet training as:

\begin{equation}
\label{eq:loss_supernet}
\begin{aligned}
    \min_{\mathbf{m}_e, \mathbf{E}, \mathbf{W}} & \mathbb{E}_{\mathbf{m}_d \sim \text{Uniform}(\mathcal{S}_e)} \mathcal{L}_{\text{CE}}(\mathcal{D}|\{\hat{\mathbf{E}}, \mathbf{W}\}) + \alpha \mathcal{L}_s, \\
    & \hat{\mathbf{E}} = \mathbf{E}_s \odot \mathbf{m}_d =  \mathbf{E} \odot  \mathbf{m}_e \odot \mathbf{m}_d  .
\end{aligned}
\end{equation}


\subsubsection{Search Strategy} 
\begin{figure}[!htbp]
    \centering
    \subfigure[Mutation]{
    \begin{minipage}[t]{0.22\textwidth}
    \centering
    \includegraphics[width=\textwidth]{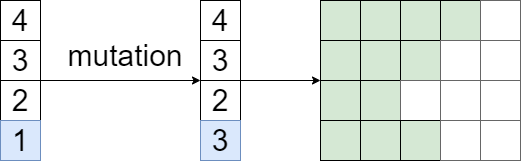}
    \label{fig:ES-mutation}
    \vspace{-5pt}
    \end{minipage}
    }
    \subfigure[Crossover]{
    \begin{minipage}[t]{0.22\textwidth}
    \centering
    \includegraphics[width=\textwidth]{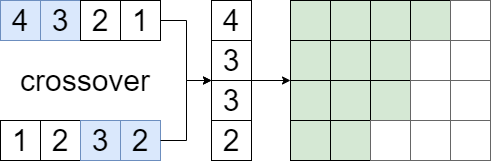}
    \label{fig:ES-crossover}
    \vspace{-5pt}
    \end{minipage}
    }
    \vspace{-10pt}
    \caption{Operations in evolutionary search.}
    \label{fig:ES}
    \vspace{-10pt}
\end{figure}

After training the supernet $\{\mathbf{E}^{*}_s, \mathbf{W^{*}_s}\}$ from Eq. \ref{eq:loss_supernet}, we present an evolutionary search for the optimal dimension mask $\mathbf{m}^{*}_d$. In the beginning, all candidates are randomly generated. At every epoch, each candidate dimension mask $\mathbf{m}_d$ is evaluated on the validation set $\mathcal{D}_{val}$ by inheriting parameters from the supernet. This part is relatively efficient as no training is involved. After the evaluation, the Top-k candidates are preserved for crossover and mutation operation to generate the candidates for the next epoch. For crossover, two randomly selected candidates are crossed to produce a new one by selecting a random point where the parents’ parts exchange happens. Fig. \ref{fig:ES-crossover} details an example where the blue parts of the two candidates are crossed. For mutation, a randomly selected candidate mutates its choice at each position with the given mutation probability $prob$. An example is illustrated in Fig. \ref{fig:ES-mutation} where the blue point is a random mutation. Crossover and mutation are repeated to generate enough new candidates given the corresponding number $n_c$ and $n_m$. After $T$ epoch, we output the best-performed candidate dimension mask as the optimal dimension mask $\mathbf{m}_d^*$. This process is shown in Algorithm \ref{alg:optembed} in line 7-17. \vspace{-10pt}

\begin{algorithm}
	\caption{The OptEmbed Algorithm} 
    \label{alg:optembed}
	\begin{algorithmic}[1]
		\Require training dataset $\mathcal{D}$, validation dataset $\mathcal{D}_{val}$
        \Ensure optimal embedding table $\mathbf{E}^*$ and model parameters $\mathbf{W}^*$
        
        \State \textbf{\#\# Supernet Training and Embedding Pruning \#\#}
        \While {not converge}
            \State Sample a mini-batch from the training dataset
            \State$\{\hat{\mathbf{E}_s},\hat{\mathbf{W}_s}\}$, $\mathbf{m}_e$ = SupernetTrain($\mathbf{\mathcal{D}}$) \Comment{Eq. \ref{eq:loss_supernet}}
        \EndWhile 
        \State $\mathbf{m}_e^*$ = GetBestPerform($\{\mathbf{m}_e\}$)
        
        \State \textbf{\#\# Dimension Mask Searching \#\#}
        \State $\tau = 0$; $P_{\tau}$ = Initialize\_population($n_m + n_c$); Topk = $\emptyset$;
        \While {$\tau < T$}
            \State $\text{AUC}_{\tau}$ = Inference($\hat{\mathbf{E}_s}$, $\hat{\mathbf{W}_s}$, $\mathcal{D}_{val}$, $P_{\tau}$);
            \State Topk = Update\_Topk(Topk, $P_{\tau}$, $\text{AUC}_{\tau}$);
            \State $P^c_{\tau}$ = Crossover(Topk, $n_c$);
            \State $P^m_{\tau}$ = Mutation(Topk, $n_m$, $prob$);
            \State $P_{\tau+1} = P^m_{\tau} \cup P^c_{\tau}$;
            \State $\tau = \tau + 1$;
        \EndWhile
        \State $\mathbf{m}_d^*$ = GetBestCand($P_{\tau}$) \Comment{Eq. \ref{eq:one-shot}}
        
        \State \textbf{\#\# Re-training\#\#}

        \State Retrain $\{\mathbf{E}^*,\mathbf{W}^*\}$ given $\mathbf{m}_e^{*}$ and $\mathbf{m}_d^{*}$ \Comment{Eq. \ref{eq:retrain}}
        
	\end{algorithmic}
\end{algorithm}
\vspace{-10pt}

\subsection{Parameter Re-training}
\label{sec:retrain}

During the supernet training, we map all raw features into embeddings. Thus to eliminate the influence of these embeddings, a re-training stage is desired to train the model with only optimal number of embeddings and embedding dimensions. The embedding mask $\mathbf{m}_e^{*}$ and the field-wise dimension mask $\mathbf{m}_d^{*}$ are obtained followed Eq. \ref{eq:loss_supernet}. In the re-training stage, the objective becomes
\begin{equation}
\begin{aligned}
    & \text{argmin}_{\mathbf{E},\mathbf{W}} \mathcal{L}_{\text{CE}}(\mathcal{D} | \{ \mathbf{E} \odot \mathbf{m}_e^* \odot \mathbf{m}_d^*, \mathbf{W}\}). \\
\end{aligned}
\label{eq:retrain}
\end{equation}
In summary, the overall process of OptEmbed can be summarized as Algorithm \ref{alg:optembed}.


\subsection{Method Discussion}
\label{sec:summary}

The combination of pruning redundant embeddings and embedding dimension search makes our OptEmbed approach efficient and effective. Table \ref{Table:summary} performs a comprehensive comparison of our approach with others that optimize embedding table on whether they satisfied the Req. \ref{def:r1}, \ref{def:r2} and \ref{def:r3}. Some methods~\cite{MDE,DNIS,AutoDim,AutoIAS,MGQE} search optimal dimension for embedding table from different granularity: feature-wise (usually grouped by feature value frequency) or field-wise. Another method ~\cite{QR} uses hashing technique to reduce the number of embeddings in the embedding table. The other methods~\cite{UMEC,PEP} utilize pruning techniques ~\cite{DST,LTH} to learn the sparse embedding table directly, which is hard to compatible with the hardware. OptEmbed method is the only method that satisfies Req. \ref{def:r1}, \ref{def:r2} and \ref{def:r3}. The rest of the method tends to violate one or two requirements.

\begin{table}[htbp]   
\renewcommand\arraystretch{0.9}
\centering
\vspace{-5pt}
\caption{Comparison of embedding learning approaches.}  
\vspace{-5pt}
\begin{tabular}{c|c|c|c} \hline    
Approach                    & R1: N.R.F.    & R2: E.D.F.    & R3: H.F.      \\ \hline
MDE~\cite{MDE}              & \XSolidBrush  & \Checkmark    & \Checkmark    \\ \hline
DNIS~\cite{DNIS}            & \XSolidBrush  & \Checkmark    & \Checkmark    \\ \hline
AutoDim~\cite{AutoDim}      & \XSolidBrush  & \Checkmark    & \Checkmark    \\ \hline
AutoField~\cite{autofield}  & \Checkmark    & \XSolidBrush  & \Checkmark    \\ \hline
QR~\cite{QR}                & \Checkmark    & \XSolidBrush  & \Checkmark    \\ \hline
PEP~\cite{PEP}              & \Checkmark    & \Checkmark    & \XSolidBrush  \\ \hline
OptEmbed                    & \Checkmark    & \Checkmark    & \Checkmark    \\ \hline
\end{tabular}
\begin{tablenotes}
\footnotesize
\item[1] \textit{N.R.F}, \textit{E.D.F.} and \textit{H.F.} are abbreviations for No Redundant Feature, Embedding Dimension Flexible and Hardwar Friendly.
\vspace{-10pt}
\end{tablenotes}
\label{Table:summary}
\end{table}
\section{experiment}
\label{sec:experiment}

We design experiments to answer the following research questions: 

\begin{itemize}[noitemsep,nolistsep,topsep=0pt,leftmargin=*]
    \item \textbf{RQ1}: Could OptEmbed achieve superior performance compared with mainstream CTR prediction models and other algorithms that optimize the embedding table?
    \item \textbf{RQ2}: How does each component of OptEmbed contribute to the final result?
    \item \textbf{RQ3}: What is the impact of the re-training stage in OptEmbed on the final result?
    \item \textbf{RQ4}: How efficient is OptEmbed compared with SOTA handcrafted models and other algorithms that optimize the embedding table?
    \item \textbf{RQ5}: Does OptEmbed output the optimal embedding table? 
\end{itemize}

\subsection{Experiment Setup}
\subsubsection{Datasets}

We conduct our experiments on three public datasets. In all following dataset, we randomly split them into $8:1:1$ as the training set, validation set, and test set respectively. 

\textbf{Criteo}\footnote{https://www.kaggle.com/c/criteo-display-ad-challenge} dataset consists of ad click data over a week. It consists of 26 categorical feature fields and 13 numerical feature fields. We follow the winner solution of the Criteo contest to discretize each numeric value $x$ to $\lfloor\log^2(x)\rfloor$, if $x>2$; $x=1$ otherwise. Following the best practice~\cite{fuxictr}, we replace infrequent categorical features with a default "OOV" (i.e. out-of-vocabulary) token, with min\_count=2.

\textbf{Avazu}\footnote{http://www.kaggle.com/c/avazu-ctr-prediction} dataset contains 10 days of click logs. It has 24 fields with categorical features, including instance id, app id, device id, etc. Following the best practice~\cite{fuxictr}, we remove the instance id field and transform the timestamp field into three new fields: hour, weekday and is\_weekend. We replace infrequent categorical features with the "OOV" token, with min\_count=2.

\textbf{KDD12}\footnote{http://www.kddcup2012.org/c/kddcup2012-track2/data} dataset contains training instances derived from search session logs. It has 11 categorical fields, and the click field is the number of times the user clicks the ad. We replace infrequent features with an "OOV" token, with min\_count=10.


\begin{table*}[!htbp]
\renewcommand\arraystretch{0.95}
\centering
\vspace{-5pt}
\caption{Overall Performance Comparison.}	\label{Table:overall}
\vspace{-5pt}
\resizebox{.98\textwidth}{!}{
\begin{tabular}{c|c|ccc|ccc|ccc|ccc}
    \hline
        & \multirow{2}{*}{Dataset} & \multicolumn{3}{c}{DeepFM} & \multicolumn{3}{|c}{DCN} & \multicolumn{3}{|c}{FNN} & \multicolumn{3}{|c}{IPNN} \\
    \cline{3-14}
        & 
        & AUC & Logloss & Sparsity & AUC & Logloss & Sparsity & AUC & Logloss & Sparsity & AUC & Logloss & Sparsity \\
    \hline
        \multirow{6}{*}{\rotatebox{90}{Criteo}}
            & Original  & 0.8104 & 0.4409 & -      & 0.8106 & 0.4408 & -      & 0.8110 & 0.4404 & -      & 0.8113 & 0.4401 & -      \\
            \cline{2-14}
            & AutoDim   & 0.8093 & 0.4420 & 0.8642 & 0.8096 & 0.4418 & 0.7917 & 0.8104 & 0.4410 & $\textbf{0.7187}$ & 0.8103 & 0.4411 & $\textbf{0.7179}$ \\
            & AutoField & 0.8101 & 0.4412 & 0.0009 & 0.8108 & 0.4405 & 0.4108 & 0.8108 & 0.4406 & 0.6221 & 0.8111 & 0.4403 & 0.3941 \\  
            & QR        & 0.8084 & 0.4444 & 0.5000 & 0.8103 & 0.4411 & 0.5000 & 0.8105 & 0.4408 & 0.5000 & 0.8102 & 0.4411 & 0.5000 \\
            & PEP       & 0.7980 & 0.4541 & 0.5010 & 0.8110 & 0.4404 & 0.5802 & 0.8108 & 0.4406 & 0.5802 & 0.8111 & 0.4402 & 0.5607 \\
            & OptEmbed  & 
            $\textbf{0.8105}$ & $\textbf{0.4409}$ & $\textbf{0.9684}$ & $\textbf{0.8113}$ & $\textbf{0.4402}$ & $\textbf{0.8534}$ & $\textbf{0.8114}$ & $\textbf{0.4400}$ & 0.6710 & $\textbf{0.8114}$ & $\textbf{0.4401}$ & 0.7122 \\
    \hline
        \multirow{6}{*}{\rotatebox{90}{Avazu}}
            & Original  & 0.7884 & 0.3751 & -      & 0.7894 & 0.3748 & -      & 0.7896 & 0.3748 & -      & 0.7898 & 0.3745 & -      \\
            \cline{2-14}
            & AutoDim   & 0.7843 & 0.3779 & \textbf{0.6936} & 0.7893 & 0.3744 & 0.5013 & 0.7894 & $\textbf{0.3743}$ & 0.5017 & 0.7894 & 0.3743 & 0.3892 \\
            & AutoField & 0.7866 & 0.3762 & 0.0020 & 0.7887 & 0.3748 & 0.0001 & 0.7892 & 0.3748 & 0.0001 & 0.7897 & 0.3744 & 0.0001 \\  
            & QR        & 0.7762 & 0.3821 & 0.5000 & 0.7868 & 0.3766 & 0.5000 & 0.7857 & 0.3769 & 0.5000 & 0.7849 & 0.3781 & $\textbf{0.5000}$ \\
            & PEP       & 0.7877 & 0.3754 & 0.4126 & 0.7896 & 0.3743 & 0.3016 & 0.7894 & 0.3744 & 0.3016 & 0.7897 & 0.3742 & 0.3016 \\
            & OptEmbed  & 
            $\textbf{0.7888}^*$ & $\textbf{0.3750}^*$ & $0.3927$ & $\textbf{0.7901}^*$ & $\textbf{0.3740}$ & $\textbf{0.6840}$ & $\textbf{0.7902}^*$ & 0.3744 & $\textbf{0.5563}$ & $\textbf{0.7902}$ & $\textbf{0.3740}^*$ & 0.4693 \\
    \hline
        \multirow{6}{*}{\rotatebox{90}{KDD12}}
            & Original  & 0.7962 & 0.1532 & -      & 0.8010 & 0.1522 & -      & 0.8008 & 0.1522 & -      & 0.8007 & 0.1522 & -      \\
            \cline{2-14}
            & AutoDim   & 0.7886 & 0.1550 & 0.0029 & 0.8016 & 0.1520 & 0.1904 & 0.8012 & 0.1522 & 0.1669 & 0.8013 & 0.1521 & 0.2286 \\
            & AutoField & 0.7953 & 0.1534 & 0.0038 & 0.8011 & 0.1525 & 0.0000 & 0.8006 & 0.1522 & 0.0000 & 0.8006 & 0.1522 & 0.0038 \\  
            & QR        & 0.7913 & 0.1544 & 0.5000 & 0.7925 & 0.1541 & $\textbf{0.5000}$ & 0.7938 & 0.1538 & 0.5000 & 0.7928 & 0.1540 & $\textbf{0.5000}$ \\
            & PEP       & 0.7957 & 0.1533 & 0.1001 & 0.7992 & 0.1525 & 0.1003 & 0.7984 & 0.1527 & 0.1003 & 0.7957 & 0.1535 & 0.1003 \\ 
            & OptEmbed  & 
            $\textbf{0.7971}^*$ & $\textbf{0.1530}^*$ & $\textbf{0.6183}$ 
            & $\textbf{0.8021}^*$ & $\textbf{0.1519}$ & 0.4715 
            & $\textbf{0.8027}^*$ & $\textbf{0.1522}$ & $\textbf{0.5105}$ 
            & $\textbf{0.8028}^*$ & $\textbf{0.1521}$ & 0.4154 \\
    \hline
\end{tabular}
}
\begin{tablenotes}
\footnotesize
\item[1] Here $*$ denotes statistically significant improvement (measured by a two-sided t-test with p-value $<0.05$) over the best baseline.
\end{tablenotes}
\vspace{-5pt}
\end{table*}

\subsubsection{Metrics}
Following the previous works~\cite{DeepFM,FNN,IPNN}, we adopt the commonly used evaluation metric in CTR prediction community: \textbf{AUC} (Area Under ROC) and \textbf{Log loss} (cross-entropy). Notes that in CTR prediction task, $\textbf{0.1 \%}$ AUC improvement is considered significant~\cite{AutoDim,autofield}. Besides, we also record the \textbf{sparsity} ratio of the embedding table, the \textbf{inference time} per batch and the \textbf{training time} of models to measure efficiency. The \textbf{sparsity} ratio is calculated as follows:

\begin{equation}
\label{eq:sparsity}
    \text{Sparsity} = 1 - \frac{\text{\#Remaining Params}}{ |f| \times D }.
\end{equation}

\subsubsection{Baseline Models}

We compare the proposed method OptEmbed with the following embedding architecture search methods: 
(i) AutoDim~\cite{AutoDim}: This baseline utilizes neural architecture search techniques\cite{DARTS} to select feasible embedding dimensions from a set of pre-defined search space. 
(ii) AutoField~\cite{autofield}: This baseline utilizes neural architecture search techniques~\cite{DARTS} to select the essential feature fields. 
(iii) QR~\cite{QR}: This baseline utilizes the Quotient-Remainder hashing trick to reduce the number of features explicitly.
(iv) PEP~\cite{PEP}: This baseline adopts trainable thresholds to remove redundant elements in the embedding table. 
We apply the above baselines and OptEmbed method over the following well-known models: DeepFM~\cite{DeepFM}, DCN~\cite{DCN}, FNN~\cite{FNN}, and IPNN~\cite{IPNN}.

\subsubsection{Implementation Details}
In this section, we provide the implementation details. For OptEmbed, (i) General hyper-params: We set the embedding dimension as 64 and batch size as 2048. For the MLP layer, we use three fully-connected layers of size [1024, 512, 256]. Following previous work~\cite{IPNN}, Adam optimizer, Batch Normalization~\cite{BatchNorm} and Xavier initialization~\cite{Xavier} are adopted. We select the optimal learning ratio from \{1e-3, 3e-4, 1e-4, 3e-5, 1e-5\} and $l_2$ regularization from \{1e-3, 3e-4, 1e-4, 3e-5, 1e-5, 3e-6, 1e-6\}. (ii) feature mask hyper-params: we select the optimal threshold learning ratio $\text{lr}_\text{t}$ from \{1e-2, 1e-3, 1e-4\} and threshold regularization $\alpha$ from \{1e-4, 3e-5, 1e-5, 3e-6, 1e-6\}.  (iii) embedding mask hyper-params: we adopt the same hyper-parameters from previous work\cite{One-shot}. For all the dimension search experiments, we empirically set mutation number $n_m = 10$, crossover number $n_c = 10$, max iteration $T=30$ and mutation probability $prob = 0.1$. During the re-training phase, we reuse the optimal learning ratio and $l_2$ regularization. For AutoDim, AutoField and PEP, we select the optimal hyper-parameter from the same hyper-parameter domain of OptEmbed.

Our implementation\footnote{https://github.com/fuyuanlyu/OptEmbed} is based on a public Pytorch library for CTR prediction\footnote{https://github.com/rixwew/pytorch-fm}. For other comparison methods, we reuse the official implementation for the PEP\footnote{https://github.com/ssui-liu/learnable-embed-sizes-for-RecSys}\cite{PEP} and QR\footnote{https://github.com/facebookresearch/dlrm}\cite{QR} methods. Due to the lack of available implementation for the AutoDim\cite{AutoDim} and AutoField\cite{autofield} method, we re-implement them based on the details provided by the authors.




\subsection{Overall Performace (RQ1)}

The overall performance of our OptEmbed and other baselines on four different models using three datasets are reported in Table \ref{Table:overall}. We summarize our observations below.

First, our OptEmbed is effective and efficient compared with the original model and other baselines. OptEmbed can achieve higher AUC with fewer parameters. However, the benefit brought by OptEmbed differs on various datasets. On Criteo, the benefit tends to be memory reduction. OptEmbed is able to reduce 67\% $\sim$ 97\% parameters with improvement not considered sigificant statistically. On KDD12 and Avazu datasets, the benefit tends to be both performance boosting and memory reduction. OptEmbed can significantly increase the AUC by up to 0.15\% compared with the original model while saving roughly $\sim$50\% of the parameters.

Secondly, among all baselines, PEP is the most similar to OptEmbed. It also tends to be the best-performed baseline on Criteo and Avazu datasets. However, it might be surpassed by AutoField and AutoDim on KDD12. Its inconsistency highlights the necessity of OptEmbed framework. Moreover, the searching phase of PEP will only stop once the embedding table reaches a predetermined sparsity ratio, completely neglecting the model performance. Such stopping criteria may result in a sub-optimal embedding table.

Finally, other baselines tend to behave differently under different cases. Without considering the effect of redundant features, \textit{AutoDim} performs well under certain cases but may result in a significant performance decrease sometime. On the other hand, \textit{AutoField} often results in low sparsity as its granularity is too large. The performance degrade brought by \textit{QR} is usually higher than other baselines. This might be related to its hashing trick, as it blindly forces different features to merge into one without considering the performance of its embedding.

\subsection{Ablation on Different Components(RQ2)}

\begin{table}[!htbp]
\renewcommand\arraystretch{0.95}
\centering
\vspace{-5pt}
\caption{Performance Comparison for Component Analysis.}
\vspace{-5pt}
\label{Table:component}
\resizebox{.48\textwidth}{!}{
\begin{tabular}{c|c|c|ccc}
    \hline
        & Basic & \multirow{2}{*}{Metrics} & \multicolumn{3}{c}{Metrics} \\
    \cline{4-6}
        & Model & & AUC & Logloss & Sparsity \\
    \hline
        \multirow{8}{*}{\rotatebox{90}{Criteo}} & \multirow{4}{*}{DeepFM} 
        &   Original    & 0.8104 & 0.4409 & -      \\
        & & OptEmbed-E  & 0.8104 & 0.4410 & 0.6267 \\
        & & OptEmbed-D  & 0.8103 & 0.4410 & 0.5547 \\
        & & OptEmbed    & 0.8105 & 0.4409 & 0.9684 \\
    \cline{2-6}
        & \multirow{4}{*}{DCN} 
        &   Original    & 0.8106 & 0.4408 & -      \\
        & & OptEmbed-E  & 0.8110 & 0.4404 & 0.6111 \\
        & & OptEmbed-D  & 0.8110 & 0.4403 & 0.7192 \\
        & & OptEmbed    & 0.8113 & 0.4402 & 0.8534 \\ 
    \hline
        \multirow{8}{*}{\rotatebox{90}{Avazu}} & \multirow{4}{*}{DeepFM} 
        &   Original    & 0.7884 & 0.3751 & -      \\
        & & OptEmbed-E  & 0.7884 & 0.3752 & 0.0000 \\
        & & OptEmbed-D  & 0.7888 & 0.3750 & 0.3927 \\
        & & OptEmbed    & 0.7888 & 0.3750 & 0.3927 \\
    \cline{2-6}
        & \multirow{4}{*}{DCN} 
        &   Original    & 0.7894 & 0.3748 & -      \\
        & & OptEmbed-E  & 0.7895 & 0.3746 & 0.0024 \\
        & & OptEmbed-D  & 0.7900 & 0.3740 & 0.5044 \\
        & & OptEmbed    & 0.7900 & 0.3743 & 0.6840 \\
    \hline
\end{tabular}
}
\vspace{-10pt}
\end{table}
In this section, we discuss the influence of different components of OptEmbed. Here we adopt two variants of OptEmbed: OptEmbed-E for only using the embedding pruning component and OptEmbed-D for only using the dimension search component. The results are shown in Table \ref{Table:component}. As we can observe, embedding pruning and dimension search components behave differently given various datasets. On the Criteo dataset, both the components reduce the embedding parameters. On DCN model, OptEmbed-E and OptEmbed-D can slightly improve model performance. OptEmbed combines these two components to obtain an optimal embedding table with fewer parameters and higher model performance. On the Avazu dataset, OptEmbed-E makes no significant difference compared with original model. This may be due to the overwhelming majority of feature values in the Avazu dataset being ID features, which tend to be informative in prediction. Hence, the optimal embedding table obtained by OptEmbed and OptEmbed-D usually is similar to each other. In all, these two components should be utilized in a unifying way to obtain the optimal embedding table considering differences between datasets.

\subsection{Ablation on Re-training(RQ3)}
We investigate the necessity of Section \ref{sec:retrain} upon the result of the DCN model over both Criteo and Avazu datasets. Results are shown in Table \ref{Table:retrain}. We compare the performance of OptEmbed under different settings with and without re-training. It can be observed that re-training can generally improve the performance. Without re-training, the neural network will inherit the sub-optimal model parameters from the supernet, which is trained for predicting the performance of all possible field-wise dimension masks. Re-training makes the model parameter optimal under the constraint of the embedding mask and field-wise dimension mask.

\begin{table}[!htbp]
    \renewcommand\arraystretch{0.95}
	\centering
	\vspace{-5pt}
	\caption{Ablation About Re-training Stage.}
	\vspace{-5pt}
	\resizebox{.49\textwidth}{!}{
	\begin{tabular}{c|cc|cc|cc}
		\hline
		  Dataset & \multicolumn{2}{c|}{Criteo} & \multicolumn{2}{c|}{Avazu} & \multicolumn{2}{c}{KDD12} \\
		\hline
		  Retrain & w. & w.o. & w. & w.o. & w. & w.o. \\
		\hline
		  AUC       & 0.8113 & 0.8110 & 0.7900 & 0.7895 & 0.8021 & 0.8005 \\
		  Logloss   & 0.4402 & 0.4404 & 0.3743 & 0.3749 & 0.1523 & 0.1526 \\
		\hline
	\end{tabular}
	}
	\begin{tablenotes}
    \footnotesize
    \item[1] \textit{w.} stands for with re-training. \textit{w.o.} stands for without re-training.
    \vspace{-10pt}
    \end{tablenotes}
	\label{Table:retrain}
\end{table}

\subsection{Efficiency Analysis(RQ4)}

In addition to the model effectiveness, training and inference efficiency are also vital when deploying the CTR prediction model into reality. In this section, we investigate the efficiency of OptEmbed from both the time and space aspects.

\subsubsection{Time Complexity}
\begin{figure}[!htbp]
    \centering
    \subfigure[Total Training Time (h)]{
    \begin{minipage}[t]{0.22\textwidth}
    \centering
    \includegraphics[width=\textwidth]{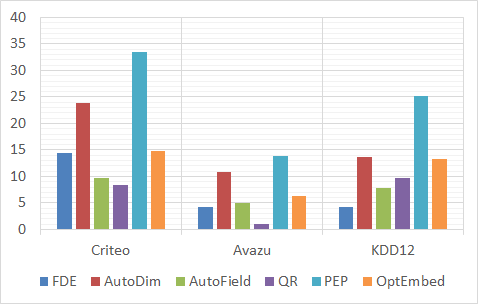}
    \vspace{-5pt}
    \label{fig:Time-Training}
    \end{minipage}
    }
    \subfigure[Inference Time (ms)]{
    \begin{minipage}[t]{0.22\textwidth}
    \centering
    \includegraphics[width=\textwidth]{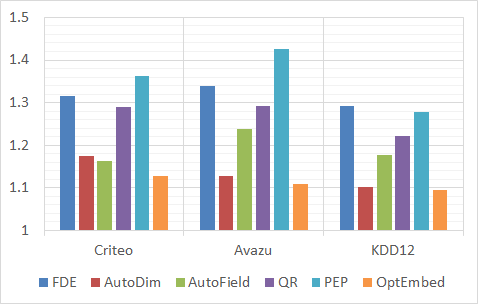}
    \vspace{-5pt}
    \label{fig:Time-Inference}
    \end{minipage}
    }
    \vspace{-10pt}
    \caption{A case study about the candidate setting.}
    \vspace{-10pt}
    \label{fig:Time}
\end{figure}

We illustrate the total training and inference time of DeepFM model trained on all three datasets in Fig. \ref{fig:Time}. Here we define the total training time as the sum of mask searching time(the time required to obtain the embedding and/or dimension mask given different methods) and re-training time(the time for re-training the parameters under the constraint of the embedding and/or dimension mask). 

For the total training time in Fig. \ref{fig:Time-Training}, we can observe that QR and AutoField tend to have faster speeds than other methods. For QR, no network architecture search is involved. It also has a smaller embedding table, leading to a faster training speed per epoch. AutoField has a smaller search space than other baselines since it only contains the feature field. Surprisingly, the total training time of original model is not always the fastest. This is because original model may take more epochs to converge. OptEmbed is faster than PEP and AutoDim because they have respectively slower convergence speeds during the mask searching and re-training phase. 

The inference time is crucial when deploying the model in reality. As shown in Fig. \ref{fig:Time-Inference}, OptEmbed achieves the least inference time. This is because the final embedding table obtained by OptEmbed has the least parameters. PEP requires the longest inference time, even longer than original model, because its embedding table tends to be sparse and hardware-unfriendly. 
Note that it is inevitable to cost additional time to search masks for OptEmbed. However, the cost is worth considering the performance increase and inference time saving, which are more important in practice.

\subsubsection{Space Complexity}
\begin{figure}[!htbp]
    \centering
    \subfigure[Criteo]{
    \begin{minipage}[t]{0.2\textwidth}
    \centering
    \includegraphics[width=\textwidth]{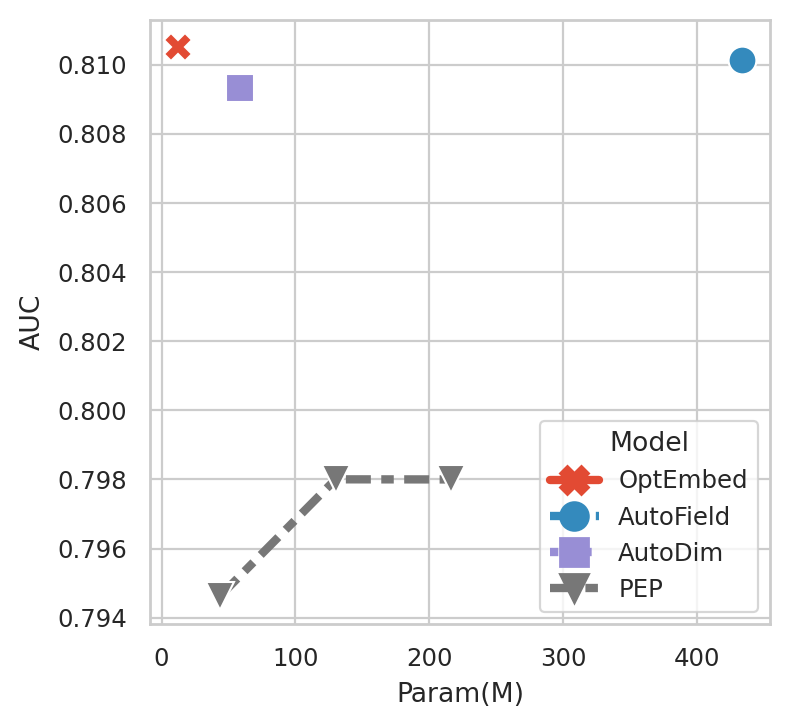}
    \label{fig:Param-AUC-Criteo}
    \vspace{-5pt}
    \end{minipage}
    }
    \subfigure[KDD12]{
    \begin{minipage}[t]{0.2\textwidth}
    \centering
    \includegraphics[width=\textwidth]{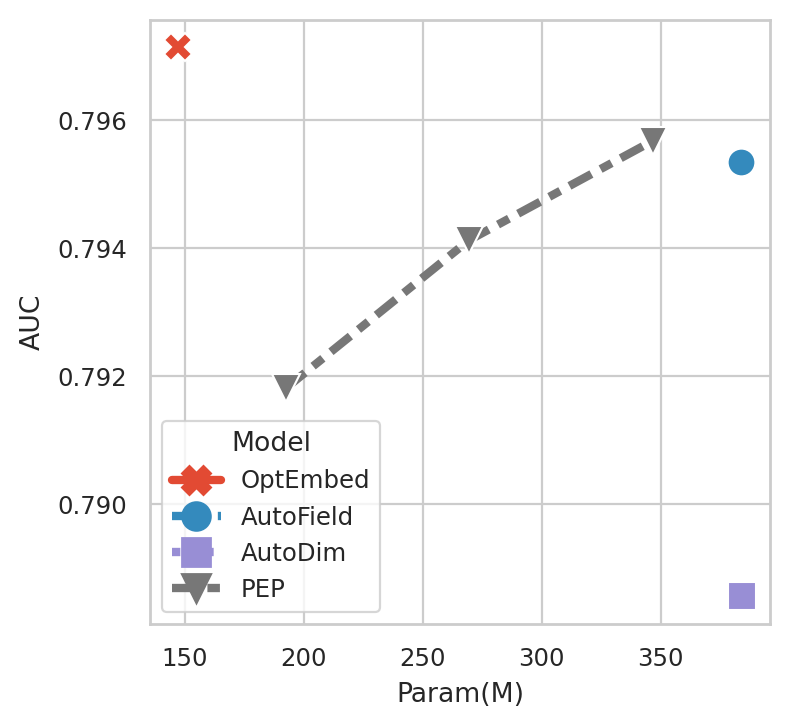}
    \label{fig:Param-AUC-KDD12}
    \vspace{-5pt}
    \end{minipage}
    }
    \vspace{-10pt}
    \caption{Visualization of efficiency-effectiveness trade-off for different datasets. The closer to the top-left the better.}
    \label{fig:Param-AUC}
    \vspace{-5pt}
\end{figure}

We plot the parameter-AUC curve of the DeepFM model on both Criteo and KDD12 datasets in Fig. \ref{fig:Param-AUC}, which reflects the relationship between the space complexity of the embedding table and model performance. There are multiple PEP points as it requires predetermined sparsity ratios as stopping criteria. So we can easily control the final sparsity ratio. AutoDim, AutoField and OptEmbed primarily aim to improve model performance. However, there is no guarantee of the final sparsity ratio. Hence we only plot one point for each method.
From Fig. \ref{fig:Param-AUC} we can make the following observations: (i) OptEmbed outperforms other baselines with the highest AUC score and the smallest embedding size. (ii) Model performance of PEP tends to degrade with the decrease of embedding parameters. (iii) AutoDim and AutoField only optimize the embedding table along one axis. Hence they do not have stable performance among datasets. They perform well on the Criteo dataset. However, they are surpassed by PEP on the KDD12 dataset.

\subsection{Case Study(RQ5)}
\begin{figure}[!htbp]
    \centering
    \includegraphics[width=0.4\textwidth]{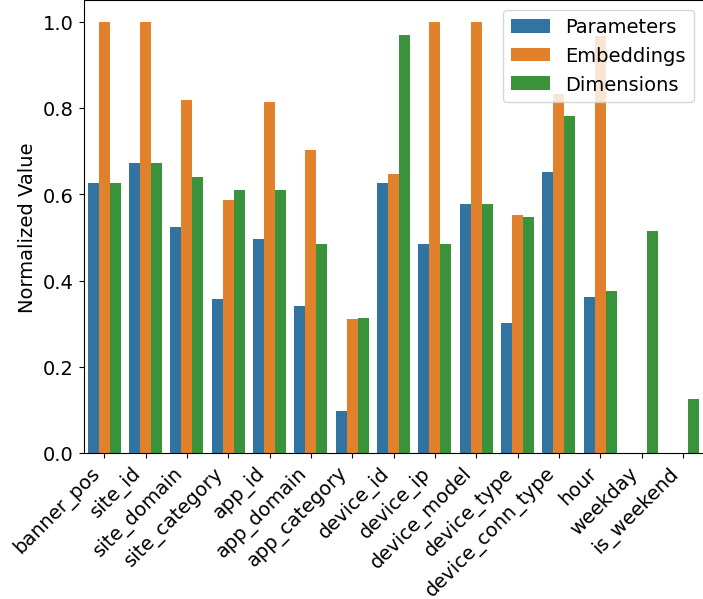}
    \vspace{-10pt}
    \caption{Case Study of OptEmbed output on Avazu}
    \label{fig:case}
    \vspace{-10pt}
\end{figure}

This section uses a case study to investigate the optimal embedding table obtained from OptEmbed. We select the embedding table of FNN model trained on Avazu dataset as an example and exclude all anonymous feature fields. Criteo and KDD12 datasets are not selected because all fields are anonymous. In Fig. \ref{fig:case}, we plot total parameters, embedding numbers, dimensions and normalize them with corresponding values from original model, respectively. We can make the following observations. First, each field's optimal dimensions and remaining feature values vary from one to another. This highlights the necessity for OptEmbed. Second, id-like features (like \textit{site\_id}, \textit{app\_id}, \textit{device\_id}) tend to have higher values than others. Such an observation is consistent with human intuition as the id-like features are the core of collaborative filtering-based recommender system. Third, it is surprising to find out that no embedding is assigned for the \textit{weekday} and \textit{is\_weekday} fields. These two fields are manually created following the best practice~\cite{fuxictr}. However, their contained information is more likely to be covered by the $\textit{hour}$ field. Such an observation shows the limit of human-defined feature selection methods.

\section{related work}
\label{sec:rw}

We discuss how our work is situated in two research topics: CTR prediction and embedding table optimization. Many machine learning models have been developed for CTR prediction ~\cite{ADS,LR,FM}. Due to the powerful learning ability, the mainstream CTR prediction research is dominated by deep learning models~\cite{fuxictr,DLRM}. Wide\&Deep~\cite{Wide_Deep} and FNN~\cite{FNN} introduce an embedding table to transform the raw inputs and an MLP layer to model high-level representations. DeepFM~\cite{DeepFM}, DCN~\cite{DCN} and IPNN~\cite{IPNN} rely on various feature interaction layers to improve performance. Increasing research focus on how to model complex feature interaction~\cite{AutoFis, AutoFeature,optinter}. With AutoML technique, AutoFIS~\cite{AutoFis} and AutoFeature~\cite{AutoFeature} search for feature interaction instead of modeling with implicit layer. 
OptEmbed is perpendicular to all these researches by providing an optimal embedidng table for various base models.

Studies on optimizing embedding tables can mainly be categorized into reducing memory usage, searching embedding dimensions, and pruning redundant values. Hash embedding~\cite{featurehashing} designs a hash method to reduce the embedding table size. Double hash~\cite{doublehashing} adopts the double hash method for one feature value to reduce the collision in hashing method. Q-R trick~\cite{QR} is also introduced to conquer the collision problem. Some quantized techniques~\cite{MGQE,xlightfm} are also borrowed for compressing the embedding table. To search field-wise embedding dimension, NAS has been utilized automatically based on well-defined search space~\cite{AutoDim,MDE, DNIS}. NAS has also been utilized to search informative feature field~\cite{autofield} automatically. AutoIAS~\cite{AutoIAS} introduces a one-shot search for both embedding dimension and architecture. Pruning redundant values in embedding tables attracted more attention recently. PEP~\cite{PEP} designs a soft threshold method to filter out low magnitude values in the embedding table with a predetermined sparsity ratio. DeepLight~\cite{deeplight} prune embedding table and other components on a pre-train network. The single-shot~\cite{single-shot} pruning method is also used to prune the embedding table carefully. This paper proposes three requirements for an optimal embedding table in Req. \ref{def:r1}, \ref{def:r2} and \ref{def:r3}, and make a thorough comparison with these works. To the best of our knowledge, we are the first to design an optimal embedding framework satisfying three requirements for CTR prediction. 
\section{conclusion}
\label{sec:conclusion}

This paper first proposes the requirements for an optimal embedding table. Based on these requirements, a novel, model-agnostic framework OptEmbed is proposed. OptEmbed optimizes the embedding table in a unifying way. It is capable of combining the supernet parameter training with redundant embedding pruning. A one-shot embedding search method is proposed based on the supernet to efficiently find optimal dimensions for different fields and obtain the optimal embedding table. Extensive experiments on three large-scale datasets demonstrate the superiority of OptEmbed in terms of both model performance and model size reduction. Several ablation studies demonstrate that basic models require optimal embedding tables on various datasets. Moreover, we also interpret the obtained result on feature fields, highlighting that our method learns the optimal embedding table.

\begin{acks} 
We specifically want to thanks Dr. Jin Guo for her helpful suggestions regarding the paper writing. 
\end{acks}

\clearpage
\normalem
\bibliographystyle{ACM-Reference-Format}
\bibliography{main.bib}


\begin{thebibliography}{44}


\ifx \showCODEN    \undefined \def \showCODEN     #1{\unskip}     \fi
\ifx \showDOI      \undefined \def \showDOI       #1{#1}\fi
\ifx \showISBNx    \undefined \def \showISBNx     #1{\unskip}     \fi
\ifx \showISBNxiii \undefined \def \showISBNxiii  #1{\unskip}     \fi
\ifx \showISSN     \undefined \def \showISSN      #1{\unskip}     \fi
\ifx \showLCCN     \undefined \def \showLCCN      #1{\unskip}     \fi
\ifx \shownote     \undefined \def \shownote      #1{#1}          \fi
\ifx \showarticletitle \undefined \def \showarticletitle #1{#1}   \fi
\ifx \showURL      \undefined \def \showURL       {\relax}        \fi
\providecommand\bibfield[2]{#2}
\providecommand\bibinfo[2]{#2}
\providecommand\natexlab[1]{#1}
\providecommand\showeprint[2][]{arXiv:#2}

\bibitem[\protect\citeauthoryear{Bender, Kindermans, Zoph, Vasudevan, and
  Le}{Bender et~al\mbox{.}}{2018}]%
        {one-shotnas}
\bibfield{author}{\bibinfo{person}{Gabriel Bender}, \bibinfo{person}{Pieter-Jan
  Kindermans}, \bibinfo{person}{Barret Zoph}, \bibinfo{person}{Vijay
  Vasudevan}, {and} \bibinfo{person}{Quoc Le}.}
  \bibinfo{year}{2018}\natexlab{}.
\newblock \showarticletitle{Understanding and Simplifying One-Shot Architecture
  Search}. In \bibinfo{booktitle}{\emph{Proceedings of the 35th International
  Conference on Machine Learning}} \emph{(\bibinfo{series}{Proceedings of
  Machine Learning Research}, Vol.~\bibinfo{volume}{80})},
  \bibfield{editor}{\bibinfo{person}{Jennifer Dy} {and}
  \bibinfo{person}{Andreas Krause}} (Eds.). \bibinfo{publisher}{PMLR},
  \bibinfo{address}{Stockholmsm{\"{a}}ssan, Stockholm, Sweden},
  \bibinfo{pages}{550--559}.
\newblock


\bibitem[\protect\citeauthoryear{Cai, Zhu, and Han}{Cai et~al\mbox{.}}{2019}]%
        {proxylessnas}
\bibfield{author}{\bibinfo{person}{Han Cai}, \bibinfo{person}{Ligeng Zhu},
  {and} \bibinfo{person}{Song Han}.} \bibinfo{year}{2019}\natexlab{}.
\newblock \showarticletitle{ProxylessNAS: Direct Neural Architecture Search on
  Target Task and Hardware}. In \bibinfo{booktitle}{\emph{7th International
  Conference on Learning Representations, {ICLR} 2019}}.
  \bibinfo{publisher}{OpenReview.net}, \bibinfo{address}{New Orleans, LA, USA},
  \bibinfo{numpages}{13}~pages.
\newblock


\bibitem[\protect\citeauthoryear{Chapelle, Manavoglu, and Rosales}{Chapelle
  et~al\mbox{.}}{2015}]%
        {ADS}
\bibfield{author}{\bibinfo{person}{Olivier Chapelle}, \bibinfo{person}{Eren
  Manavoglu}, {and} \bibinfo{person}{Romer Rosales}.}
  \bibinfo{year}{2015}\natexlab{}.
\newblock \showarticletitle{Simple and Scalable Response Prediction for Display
  Advertising}.
\newblock \bibinfo{journal}{\emph{ACM Trans. Intell. Syst. Technol.}}
  \bibinfo{volume}{5}, \bibinfo{number}{4} (\bibinfo{date}{dec}
  \bibinfo{year}{2015}), \bibinfo{pages}{61}.
\newblock
\showISSN{2157-6904}


\bibitem[\protect\citeauthoryear{Cheng, Koc, Harmsen, Shaked, Chandra, Aradhye,
  Anderson, Corrado, Chai, Ispir, Anil, Haque, Hong, Jain, Liu, and Shah}{Cheng
  et~al\mbox{.}}{2016}]%
        {Wide_Deep}
\bibfield{author}{\bibinfo{person}{Heng{-}Tze Cheng}, \bibinfo{person}{Levent
  Koc}, \bibinfo{person}{Jeremiah Harmsen}, \bibinfo{person}{Tal Shaked},
  \bibinfo{person}{Tushar Chandra}, \bibinfo{person}{Hrishi Aradhye},
  \bibinfo{person}{Glen Anderson}, \bibinfo{person}{Greg Corrado},
  \bibinfo{person}{Wei Chai}, \bibinfo{person}{Mustafa Ispir},
  \bibinfo{person}{Rohan Anil}, \bibinfo{person}{Zakaria Haque},
  \bibinfo{person}{Lichan Hong}, \bibinfo{person}{Vihan Jain},
  \bibinfo{person}{Xiaobing Liu}, {and} \bibinfo{person}{Hemal Shah}.}
  \bibinfo{year}{2016}\natexlab{}.
\newblock \showarticletitle{Wide {\&} Deep Learning for Recommender Systems}.
  In \bibinfo{booktitle}{\emph{Proceedings of the 1st Workshop on Deep Learning
  for Recommender Systems, DLRS@RecSys 2016}}. \bibinfo{publisher}{{ACM}},
  \bibinfo{address}{Boston, MA, USA}, \bibinfo{pages}{7--10}.
\newblock


\bibitem[\protect\citeauthoryear{Cheng, Shen, and Huang}{Cheng
  et~al\mbox{.}}{2020}]%
        {DNIS}
\bibfield{author}{\bibinfo{person}{Weiyu Cheng}, \bibinfo{person}{Yanyan Shen},
  {and} \bibinfo{person}{Linpeng Huang}.} \bibinfo{year}{2020}\natexlab{}.
\newblock \showarticletitle{Differentiable Neural Input Search for Recommender
  Systems}.
\newblock \bibinfo{journal}{\emph{CoRR}}  \bibinfo{volume}{abs/2006.04466}
  (\bibinfo{year}{2020}).
\newblock


\bibitem[\protect\citeauthoryear{Deng, Pan, Zhou, Kong, Flores, and Lin}{Deng
  et~al\mbox{.}}{2021}]%
        {deeplight}
\bibfield{author}{\bibinfo{person}{Wei Deng}, \bibinfo{person}{Junwei Pan},
  \bibinfo{person}{Tian Zhou}, \bibinfo{person}{Deguang Kong},
  \bibinfo{person}{Aaron Flores}, {and} \bibinfo{person}{Guang Lin}.}
  \bibinfo{year}{2021}\natexlab{}.
\newblock \showarticletitle{DeepLight: Deep Lightweight Feature Interactions
  for Accelerating CTR Predictions in Ad Serving}. In
  \bibinfo{booktitle}{\emph{WSDM '21}}. \bibinfo{publisher}{{ACM}},
  \bibinfo{address}{Virtual Event, Israel}, \bibinfo{pages}{922–930}.
\newblock


\bibitem[\protect\citeauthoryear{Don{\`a} and Gallinari}{Don{\`a} and
  Gallinari}{2021}]%
        {featureselection}
\bibfield{author}{\bibinfo{person}{J{\'e}r{\'e}mie Don{\`a}} {and}
  \bibinfo{person}{Patrick Gallinari}.} \bibinfo{year}{2021}\natexlab{}.
\newblock \showarticletitle{Differentiable Feature Selection, A
  Reparameterization Approach}. In \bibinfo{booktitle}{\emph{Machine Learning
  and Knowledge Discovery in Databases. Research Track}}.
  \bibinfo{publisher}{Springer International Publishing},
  \bibinfo{address}{Spain}, \bibinfo{pages}{414--429}.
\newblock
\showISBNx{978-3-030-86523-8}


\bibitem[\protect\citeauthoryear{Frankle and Carbin}{Frankle and
  Carbin}{2019}]%
        {LTH}
\bibfield{author}{\bibinfo{person}{Jonathan Frankle} {and}
  \bibinfo{person}{Michael Carbin}.} \bibinfo{year}{2019}\natexlab{}.
\newblock \showarticletitle{The Lottery Ticket Hypothesis: Finding Sparse,
  Trainable Neural Networks}. In \bibinfo{booktitle}{\emph{7th International
  Conference on Learning Representations, {ICLR} 2019}}.
  \bibinfo{publisher}{OpenReview.net}, \bibinfo{address}{New Orleans, LA, USA},
  \bibinfo{numpages}{42}~pages.
\newblock


\bibitem[\protect\citeauthoryear{Ginart, Naumov, Mudigere, Yang, and
  Zou}{Ginart et~al\mbox{.}}{2021}]%
        {MDE}
\bibfield{author}{\bibinfo{person}{Antonio~A. Ginart}, \bibinfo{person}{Maxim
  Naumov}, \bibinfo{person}{Dheevatsa Mudigere}, \bibinfo{person}{Jiyan Yang},
  {and} \bibinfo{person}{James Zou}.} \bibinfo{year}{2021}\natexlab{}.
\newblock \showarticletitle{Mixed Dimension Embeddings with Application to
  Memory-Efficient Recommendation Systems}. In \bibinfo{booktitle}{\emph{{IEEE}
  International Symposium on Information Theory, {ISIT} 2021}}.
  \bibinfo{publisher}{{IEEE}}, \bibinfo{address}{Australia},
  \bibinfo{pages}{2786--2791}.
\newblock


\bibitem[\protect\citeauthoryear{Glorot and Bengio}{Glorot and Bengio}{2010}]%
        {Xavier}
\bibfield{author}{\bibinfo{person}{Xavier Glorot} {and} \bibinfo{person}{Yoshua
  Bengio}.} \bibinfo{year}{2010}\natexlab{}.
\newblock \showarticletitle{Understanding the difficulty of training deep
  feedforward neural networks}. In \bibinfo{booktitle}{\emph{Proceedings of the
  Thirteenth International Conference on Artificial Intelligence and
  Statistics, {AISTATS} 2010}} \emph{(\bibinfo{series}{{JMLR} Proceedings},
  Vol.~\bibinfo{volume}{9})}. \bibinfo{publisher}{JMLR.org},
  \bibinfo{address}{Italy}, \bibinfo{pages}{249--256}.
\newblock


\bibitem[\protect\citeauthoryear{Guo, Guo, Gao, Tang, He, and Liu}{Guo
  et~al\mbox{.}}{2021}]%
        {sfctr}
\bibfield{author}{\bibinfo{person}{Huifeng Guo}, \bibinfo{person}{Wei Guo},
  \bibinfo{person}{Yong Gao}, \bibinfo{person}{Ruiming Tang},
  \bibinfo{person}{Xiuqiang He}, {and} \bibinfo{person}{Wenzhi Liu}.}
  \bibinfo{year}{2021}\natexlab{}.
\newblock \bibinfo{booktitle}{\emph{ScaleFreeCTR: MixCache-Based Distributed
  Training System for CTR Models with Huge Embedding Table}}.
\newblock \bibinfo{publisher}{Association for Computing Machinery},
  \bibinfo{address}{New York, NY, USA}, \bibinfo{pages}{1269–1278}.
\newblock
\showISBNx{9781450380379}


\bibitem[\protect\citeauthoryear{Guo, Tang, Ye, Li, and He}{Guo
  et~al\mbox{.}}{2017}]%
        {DeepFM}
\bibfield{author}{\bibinfo{person}{Huifeng Guo}, \bibinfo{person}{Ruiming
  Tang}, \bibinfo{person}{Yunming Ye}, \bibinfo{person}{Zhenguo Li}, {and}
  \bibinfo{person}{Xiuqiang He}.} \bibinfo{year}{2017}\natexlab{}.
\newblock \showarticletitle{DeepFM: {A} Factorization-Machine based Neural
  Network for {CTR} Prediction}. In \bibinfo{booktitle}{\emph{Proceedings of
  the Twenty-Sixth International Joint Conference on Artificial Intelligence,
  {IJCAI} 2017}}. \bibinfo{publisher}{ijcai.org}, \bibinfo{address}{Australia},
  \bibinfo{pages}{1725--1731}.
\newblock


\bibitem[\protect\citeauthoryear{Guo, Zhang, Mu, Heng, Liu, Wei, and Sun}{Guo
  et~al\mbox{.}}{2020}]%
        {One-shot}
\bibfield{author}{\bibinfo{person}{Zichao Guo}, \bibinfo{person}{Xiangyu
  Zhang}, \bibinfo{person}{Haoyuan Mu}, \bibinfo{person}{Wen Heng},
  \bibinfo{person}{Zechun Liu}, \bibinfo{person}{Yichen Wei}, {and}
  \bibinfo{person}{Jian Sun}.} \bibinfo{year}{2020}\natexlab{}.
\newblock \showarticletitle{Single Path One-Shot Neural Architecture Search
  with Uniform Sampling}. In \bibinfo{booktitle}{\emph{Computer Vision - {ECCV}
  2020 - 16th European Conference}} \emph{(\bibinfo{series}{Lecture Notes in
  Computer Science}, Vol.~\bibinfo{volume}{12361})}.
  \bibinfo{publisher}{Springer}, \bibinfo{address}{UK},
  \bibinfo{pages}{544--560}.
\newblock


\bibitem[\protect\citeauthoryear{Han, Pool, Tran, and Dally}{Han
  et~al\mbox{.}}{2015}]%
        {weight_connect}
\bibfield{author}{\bibinfo{person}{Song Han}, \bibinfo{person}{Jeff Pool},
  \bibinfo{person}{John Tran}, {and} \bibinfo{person}{William~J. Dally}.}
  \bibinfo{year}{2015}\natexlab{}.
\newblock \showarticletitle{Learning both Weights and Connections for Efficient
  Neural Network}. In \bibinfo{booktitle}{\emph{Advances in Neural Information
  Processing Systems 28: Annual Conference on Neural Information Processing
  Systems 2015}}. \bibinfo{publisher}{Springer}, \bibinfo{address}{Canada},
  \bibinfo{pages}{1135--1143}.
\newblock


\bibitem[\protect\citeauthoryear{He and Chua}{He and Chua}{2017}]%
        {NFM}
\bibfield{author}{\bibinfo{person}{Xiangnan He} {and}
  \bibinfo{person}{Tat{-}Seng Chua}.} \bibinfo{year}{2017}\natexlab{}.
\newblock \showarticletitle{Neural Factorization Machines for Sparse Predictive
  Analytics}. In \bibinfo{booktitle}{\emph{Proceedings of the 40th
  International {ACM} {SIGIR} Conference on Research and Development in
  Information Retrieval}}, \bibfield{editor}{\bibinfo{person}{Noriko Kando},
  \bibinfo{person}{Tetsuya Sakai}, \bibinfo{person}{Hideo Joho},
  \bibinfo{person}{Hang Li}, \bibinfo{person}{Arjen~P. de~Vries}, {and}
  \bibinfo{person}{Ryen~W. White}} (Eds.). \bibinfo{publisher}{{ACM}},
  \bibinfo{address}{Shinjuku, Tokyo, Japan}, \bibinfo{pages}{355--364}.
\newblock


\bibitem[\protect\citeauthoryear{Ioffe and Szegedy}{Ioffe and Szegedy}{2015}]%
        {BatchNorm}
\bibfield{author}{\bibinfo{person}{Sergey Ioffe} {and}
  \bibinfo{person}{Christian Szegedy}.} \bibinfo{year}{2015}\natexlab{}.
\newblock \showarticletitle{Batch Normalization: Accelerating Deep Network
  Training by Reducing Internal Covariate Shift}. In
  \bibinfo{booktitle}{\emph{Proceedings of the 32nd International Conference on
  Machine Learning, {ICML} 2015}} \emph{(\bibinfo{series}{{JMLR} Workshop and
  Conference Proceedings}, Vol.~\bibinfo{volume}{37})}.
  \bibinfo{publisher}{JMLR.org}, \bibinfo{address}{France},
  \bibinfo{pages}{448--456}.
\newblock


\bibitem[\protect\citeauthoryear{Jiang, Wang, Chen, Wang, Lian, and Chen}{Jiang
  et~al\mbox{.}}{2021}]%
        {xlightfm}
\bibfield{author}{\bibinfo{person}{Gangwei Jiang}, \bibinfo{person}{Hao Wang},
  \bibinfo{person}{Jin Chen}, \bibinfo{person}{Haoyu Wang},
  \bibinfo{person}{Defu Lian}, {and} \bibinfo{person}{Enhong Chen}.}
  \bibinfo{year}{2021}\natexlab{}.
\newblock \showarticletitle{xLightFM: Extremely Memory-Efficient Factorization
  Machine}. In \bibinfo{booktitle}{\emph{{SIGIR} '21: The 44th International
  {ACM} {SIGIR} Conference on Research and Development in Information
  Retrieval}}, \bibfield{editor}{\bibinfo{person}{Fernando Diaz},
  \bibinfo{person}{Chirag Shah}, \bibinfo{person}{Torsten Suel},
  \bibinfo{person}{Pablo Castells}, \bibinfo{person}{Rosie Jones}, {and}
  \bibinfo{person}{Tetsuya Sakai}} (Eds.). \bibinfo{publisher}{{ACM}},
  \bibinfo{address}{Virtual Event, Canada}, \bibinfo{pages}{337--346}.
\newblock


\bibitem[\protect\citeauthoryear{Kang, Cheng, Chen, Yi, Lin, Hong, and
  Chi}{Kang et~al\mbox{.}}{2020}]%
        {MGQE}
\bibfield{author}{\bibinfo{person}{Wang{-}Cheng Kang},
  \bibinfo{person}{Derek~Zhiyuan Cheng}, \bibinfo{person}{Ting Chen},
  \bibinfo{person}{Xinyang Yi}, \bibinfo{person}{Dong Lin},
  \bibinfo{person}{Lichan Hong}, {and} \bibinfo{person}{Ed~H. Chi}.}
  \bibinfo{year}{2020}\natexlab{}.
\newblock \showarticletitle{Learning Multi-granular Quantized Embeddings for
  Large-Vocab Categorical Features in Recommender Systems}. In
  \bibinfo{booktitle}{\emph{Companion of The 2020 Web Conference 2020}}.
  \bibinfo{publisher}{{ACM} / {IW3C2}}, \bibinfo{address}{Taiwan},
  \bibinfo{pages}{562--566}.
\newblock


\bibitem[\protect\citeauthoryear{Khawar, Hang, Tang, Liu, Li, and He}{Khawar
  et~al\mbox{.}}{2020}]%
        {AutoFeature}
\bibfield{author}{\bibinfo{person}{Farhan Khawar}, \bibinfo{person}{Xu Hang},
  \bibinfo{person}{Ruiming Tang}, \bibinfo{person}{Bin Liu},
  \bibinfo{person}{Zhenguo Li}, {and} \bibinfo{person}{Xiuqiang He}.}
  \bibinfo{year}{2020}\natexlab{}.
\newblock \showarticletitle{AutoFeature: Searching for Feature Interactions and
  Their Architectures for Click-through Rate Prediction}. In
  \bibinfo{booktitle}{\emph{{CIKM} '20: The 29th {ACM} International Conference
  on Information and Knowledge Management}}. \bibinfo{publisher}{{ACM}},
  \bibinfo{address}{Ireland}, \bibinfo{pages}{625--634}.
\newblock


\bibitem[\protect\citeauthoryear{Liu, Zhu, Li, Zhang, Lai, Tang, He, Li, and
  Yu}{Liu et~al\mbox{.}}{2020b}]%
        {AutoFis}
\bibfield{author}{\bibinfo{person}{Bin Liu}, \bibinfo{person}{Chenxu Zhu},
  \bibinfo{person}{Guilin Li}, \bibinfo{person}{Weinan Zhang},
  \bibinfo{person}{Jincai Lai}, \bibinfo{person}{Ruiming Tang},
  \bibinfo{person}{Xiuqiang He}, \bibinfo{person}{Zhenguo Li}, {and}
  \bibinfo{person}{Yong Yu}.} \bibinfo{year}{2020}\natexlab{b}.
\newblock \showarticletitle{AutoFIS: Automatic Feature Interaction Selection in
  Factorization Models for Click-Through Rate Prediction}. In
  \bibinfo{booktitle}{\emph{{KDD} '20: The 26th {ACM} {SIGKDD} Conference on
  Knowledge Discovery and Data Mining}}. \bibinfo{publisher}{{ACM}},
  \bibinfo{address}{USA}, \bibinfo{pages}{2636--2645}.
\newblock


\bibitem[\protect\citeauthoryear{Liu, Simonyan, and Yang}{Liu
  et~al\mbox{.}}{2019}]%
        {DARTS}
\bibfield{author}{\bibinfo{person}{Hanxiao Liu}, \bibinfo{person}{Karen
  Simonyan}, {and} \bibinfo{person}{Yiming Yang}.}
  \bibinfo{year}{2019}\natexlab{}.
\newblock \showarticletitle{{DARTS:} Differentiable Architecture Search}. In
  \bibinfo{booktitle}{\emph{7th International Conference on Learning
  Representations, {ICLR} 2019}}. \bibinfo{publisher}{OpenReview.net},
  \bibinfo{address}{USA}.
\newblock


\bibitem[\protect\citeauthoryear{Liu, Xu, Shi, Cheung, and So}{Liu
  et~al\mbox{.}}{2020a}]%
        {DST}
\bibfield{author}{\bibinfo{person}{Junjie Liu}, \bibinfo{person}{Zhe Xu},
  \bibinfo{person}{Runbin Shi}, \bibinfo{person}{Ray C.~C. Cheung}, {and}
  \bibinfo{person}{Hayden~Kwok{-}Hay So}.} \bibinfo{year}{2020}\natexlab{a}.
\newblock \showarticletitle{Dynamic Sparse Training: Find Efficient Sparse
  Network From Scratch With Trainable Masked Layers}. In
  \bibinfo{booktitle}{\emph{8th International Conference on Learning
  Representations, {ICLR} 2020}}. \bibinfo{publisher}{OpenReview.net},
  \bibinfo{address}{Ethiopia}.
\newblock
\urldef\tempurl%
\url{https://openreview.net/forum?id=SJlbGJrtDB}
\showURL{%
\tempurl}


\bibitem[\protect\citeauthoryear{Liu, Gao, Chen, Jin, and Li}{Liu
  et~al\mbox{.}}{2021}]%
        {PEP}
\bibfield{author}{\bibinfo{person}{Siyi Liu}, \bibinfo{person}{Chen Gao},
  \bibinfo{person}{Yihong Chen}, \bibinfo{person}{Depeng Jin}, {and}
  \bibinfo{person}{Yong Li}.} \bibinfo{year}{2021}\natexlab{}.
\newblock \showarticletitle{Learnable Embedding sizes for Recommender Systems}.
  In \bibinfo{booktitle}{\emph{9th International Conference on Learning
  Representations, {ICLR} 2021}}. \bibinfo{publisher}{OpenReview.net},
  \bibinfo{address}{Austria}.
\newblock


\bibitem[\protect\citeauthoryear{Lyu, Tang, Guo, Tang, He, Zhang, and Liu}{Lyu
  et~al\mbox{.}}{2021}]%
        {optinter}
\bibfield{author}{\bibinfo{person}{Fuyuan Lyu}, \bibinfo{person}{Xing Tang},
  \bibinfo{person}{Huifeng Guo}, \bibinfo{person}{Ruiming Tang},
  \bibinfo{person}{Xiuqiang He}, \bibinfo{person}{Rui Zhang}, {and}
  \bibinfo{person}{Xue Liu}.} \bibinfo{year}{2021}\natexlab{}.
\newblock \showarticletitle{Memorize, Factorize, or be Na{\"{\i}}ve: Learning
  Optimal Feature Interaction Methods for {CTR} Prediction}.
\newblock \bibinfo{journal}{\emph{CoRR}}  \bibinfo{volume}{abs/2108.01265}
  (\bibinfo{year}{2021}).
\newblock
\showeprint[arXiv]{2108.01265}
\urldef\tempurl%
\url{https://arxiv.org/abs/2108.01265}
\showURL{%
\tempurl}


\bibitem[\protect\citeauthoryear{Meng, Zhang, Li, Li, Zhu, and Sun}{Meng
  et~al\mbox{.}}{2021}]%
        {AutoPI}
\bibfield{author}{\bibinfo{person}{Ze Meng}, \bibinfo{person}{Jinnian Zhang},
  \bibinfo{person}{Yumeng Li}, \bibinfo{person}{Jiancheng Li},
  \bibinfo{person}{Tanchao Zhu}, {and} \bibinfo{person}{Lifeng Sun}.}
  \bibinfo{year}{2021}\natexlab{}.
\newblock \showarticletitle{A General Method For Automatic Discovery of
  Powerful Interactions In Click-Through Rate Prediction}. In
  \bibinfo{booktitle}{\emph{{SIGIR} '21: The 44th International {ACM} {SIGIR}
  Conference on Research and Development in Information Retrieval}}.
  \bibinfo{publisher}{{ACM}}, \bibinfo{address}{Canada},
  \bibinfo{pages}{1298--1307}.
\newblock


\bibitem[\protect\citeauthoryear{Naumov, Mudigere, Shi, Huang, Sundaraman,
  Park, Wang, Gupta, Wu, Azzolini, Dzhulgakov, Mallevich, Cherniavskii, Lu,
  Krishnamoorthi, Yu, Kondratenko, Pereira, Chen, Chen, Rao, Jia, Xiong, and
  Smelyanskiy}{Naumov et~al\mbox{.}}{2019}]%
        {DLRM}
\bibfield{author}{\bibinfo{person}{Maxim Naumov}, \bibinfo{person}{Dheevatsa
  Mudigere}, \bibinfo{person}{Hao{-}Jun~Michael Shi}, \bibinfo{person}{Jianyu
  Huang}, \bibinfo{person}{Narayanan Sundaraman}, \bibinfo{person}{Jongsoo
  Park}, \bibinfo{person}{Xiaodong Wang}, \bibinfo{person}{Udit Gupta},
  \bibinfo{person}{Carole{-}Jean Wu}, \bibinfo{person}{Alisson~G. Azzolini},
  \bibinfo{person}{Dmytro Dzhulgakov}, \bibinfo{person}{Andrey Mallevich},
  \bibinfo{person}{Ilia Cherniavskii}, \bibinfo{person}{Yinghai Lu},
  \bibinfo{person}{Raghuraman Krishnamoorthi}, \bibinfo{person}{Ansha Yu},
  \bibinfo{person}{Volodymyr Kondratenko}, \bibinfo{person}{Stephanie Pereira},
  \bibinfo{person}{Xianjie Chen}, \bibinfo{person}{Wenlin Chen},
  \bibinfo{person}{Vijay Rao}, \bibinfo{person}{Bill Jia},
  \bibinfo{person}{Liang Xiong}, {and} \bibinfo{person}{Misha Smelyanskiy}.}
  \bibinfo{year}{2019}\natexlab{}.
\newblock \showarticletitle{Deep Learning Recommendation Model for
  Personalization and Recommendation Systems}.
\newblock \bibinfo{journal}{\emph{CoRR}}  \bibinfo{volume}{abs/1906.00091}
  (\bibinfo{year}{2019}).
\newblock


\bibitem[\protect\citeauthoryear{Qu, Ye, Tang, Zhang, Shi, and Yin}{Qu
  et~al\mbox{.}}{2022}]%
        {single-shot}
\bibfield{author}{\bibinfo{person}{Liang Qu}, \bibinfo{person}{Yonghong Ye},
  \bibinfo{person}{Ningzhi Tang}, \bibinfo{person}{Lixin Zhang},
  \bibinfo{person}{Yuhui Shi}, {and} \bibinfo{person}{Hongzhi Yin}.}
  \bibinfo{year}{2022}\natexlab{}.
\newblock \showarticletitle{Single-shot Embedding Dimension Search in
  Recommender System}.
\newblock \bibinfo{journal}{\emph{CoRR}}  \bibinfo{volume}{abs/2204.03281}
  (\bibinfo{year}{2022}), \bibinfo{numpages}{11}~pages.
\newblock
\urldef\tempurl%
\url{https://doi.org/10.48550/arXiv.2204.03281}
\showDOI{\tempurl}
\showeprint[arXiv]{2204.03281}


\bibitem[\protect\citeauthoryear{Qu, Fang, Zhang, Tang, Niu, Guo, Yu, and
  He}{Qu et~al\mbox{.}}{2018}]%
        {IPNN}
\bibfield{author}{\bibinfo{person}{Yanru Qu}, \bibinfo{person}{Bohui Fang},
  \bibinfo{person}{Weinan Zhang}, \bibinfo{person}{Ruiming Tang},
  \bibinfo{person}{Minzhe Niu}, \bibinfo{person}{Huifeng Guo},
  \bibinfo{person}{Yong Yu}, {and} \bibinfo{person}{Xiuqiang He}.}
  \bibinfo{year}{2018}\natexlab{}.
\newblock \showarticletitle{Product-Based Neural Networks for User Response
  Prediction over Multi-Field Categorical Data}.
\newblock \bibinfo{journal}{\emph{ACM Trans. Inf. Syst.}} \bibinfo{volume}{37},
  \bibinfo{number}{1}, Article \bibinfo{articleno}{5} (\bibinfo{date}{oct}
  \bibinfo{year}{2018}), \bibinfo{numpages}{35}~pages.
\newblock
\showISSN{1046-8188}


\bibitem[\protect\citeauthoryear{Rendle}{Rendle}{2010}]%
        {FM}
\bibfield{author}{\bibinfo{person}{Steffen Rendle}.}
  \bibinfo{year}{2010}\natexlab{}.
\newblock \showarticletitle{Factorization Machines}. In
  \bibinfo{booktitle}{\emph{{ICDM} 2010, The 10th {IEEE} International
  Conference on Data Mining}}. \bibinfo{publisher}{{IEEE} Computer Society},
  \bibinfo{address}{Australia}, \bibinfo{pages}{995--1000}.
\newblock


\bibitem[\protect\citeauthoryear{Richardson, Dominowska, and Ragno}{Richardson
  et~al\mbox{.}}{2007}]%
        {LR}
\bibfield{author}{\bibinfo{person}{Matthew Richardson}, \bibinfo{person}{Ewa
  Dominowska}, {and} \bibinfo{person}{Robert Ragno}.}
  \bibinfo{year}{2007}\natexlab{}.
\newblock \showarticletitle{Predicting Clicks: Estimating the Click-through
  Rate for New Ads}. In \bibinfo{booktitle}{\emph{Proceedings of the 16th
  International Conference on World Wide Web}} (Banff, Alberta, Canada)
  \emph{(\bibinfo{series}{WWW '07})}. \bibinfo{publisher}{Association for
  Computing Machinery}, \bibinfo{address}{New York, NY, USA},
  \bibinfo{pages}{521–530}.
\newblock
\showISBNx{9781595936547}


\bibitem[\protect\citeauthoryear{Shen, Wang, Gui, Tan, Wang, and Liu}{Shen
  et~al\mbox{.}}{2021}]%
        {UMEC}
\bibfield{author}{\bibinfo{person}{Jiayi Shen}, \bibinfo{person}{Haotao Wang},
  \bibinfo{person}{Shupeng Gui}, \bibinfo{person}{Jianchao Tan},
  \bibinfo{person}{Zhangyang Wang}, {and} \bibinfo{person}{Ji Liu}.}
  \bibinfo{year}{2021}\natexlab{}.
\newblock \showarticletitle{{UMEC:} Unified model and embedding compression for
  efficient recommendation systems}. In \bibinfo{booktitle}{\emph{9th
  International Conference on Learning Representations, {ICLR} 2021}}.
  \bibinfo{publisher}{OpenReview.net}, \bibinfo{address}{Austria}.
\newblock


\bibitem[\protect\citeauthoryear{Shi, Mudigere, Naumov, and Yang}{Shi
  et~al\mbox{.}}{2020}]%
        {QR}
\bibfield{author}{\bibinfo{person}{Hao{-}Jun~Michael Shi},
  \bibinfo{person}{Dheevatsa Mudigere}, \bibinfo{person}{Maxim Naumov}, {and}
  \bibinfo{person}{Jiyan Yang}.} \bibinfo{year}{2020}\natexlab{}.
\newblock \showarticletitle{Compositional Embeddings Using Complementary
  Partitions for Memory-Efficient Recommendation Systems}. In
  \bibinfo{booktitle}{\emph{{KDD} '20: The 26th {ACM} {SIGKDD} Conference on
  Knowledge Discovery and Data Mining}}. \bibinfo{publisher}{{ACM}},
  \bibinfo{address}{USA}, \bibinfo{pages}{165--175}.
\newblock


\bibitem[\protect\citeauthoryear{Song, Shi, Xiao, Duan, Xu, Zhang, and
  Tang}{Song et~al\mbox{.}}{2019}]%
        {AutoInt}
\bibfield{author}{\bibinfo{person}{Weiping Song}, \bibinfo{person}{Chence Shi},
  \bibinfo{person}{Zhiping Xiao}, \bibinfo{person}{Zhijian Duan},
  \bibinfo{person}{Yewen Xu}, \bibinfo{person}{Ming Zhang}, {and}
  \bibinfo{person}{Jian Tang}.} \bibinfo{year}{2019}\natexlab{}.
\newblock \showarticletitle{AutoInt: Automatic Feature Interaction Learning via
  Self-Attentive Neural Networks}. In \bibinfo{booktitle}{\emph{Proceedings of
  the 28th {ACM} International Conference on Information and Knowledge
  Management, {CIKM} 2019}}. \bibinfo{publisher}{{ACM}},
  \bibinfo{address}{China}, \bibinfo{pages}{1161--1170}.
\newblock


\bibitem[\protect\citeauthoryear{Wang, Fu, Fu, and Wang}{Wang
  et~al\mbox{.}}{2017}]%
        {DCN}
\bibfield{author}{\bibinfo{person}{Ruoxi Wang}, \bibinfo{person}{Bin Fu},
  \bibinfo{person}{Gang Fu}, {and} \bibinfo{person}{Mingliang Wang}.}
  \bibinfo{year}{2017}\natexlab{}.
\newblock \showarticletitle{Deep \& Cross Network for Ad Click Predictions}. In
  \bibinfo{booktitle}{\emph{Proceedings of the ADKDD'17}}
  \emph{(\bibinfo{series}{ADKDD'17})}. \bibinfo{publisher}{Association for
  Computing Machinery}, \bibinfo{address}{Canada}, Article
  \bibinfo{articleno}{12}, \bibinfo{numpages}{7}~pages.
\newblock


\bibitem[\protect\citeauthoryear{Wang, Zhao, Xu, and Wu}{Wang
  et~al\mbox{.}}{2022}]%
        {autofield}
\bibfield{author}{\bibinfo{person}{Yejing Wang}, \bibinfo{person}{Xiangyu
  Zhao}, \bibinfo{person}{Tong Xu}, {and} \bibinfo{person}{Xian Wu}.}
  \bibinfo{year}{2022}\natexlab{}.
\newblock \showarticletitle{AutoField: Automating Feature Selection in Deep
  Recommender Systems}. In \bibinfo{booktitle}{\emph{Proceedings of the ACM Web
  Conference 2022}} (Virtual Event, Lyon, France) \emph{(\bibinfo{series}{WWW
  '22})}. \bibinfo{publisher}{Association for Computing Machinery},
  \bibinfo{address}{New York, NY, USA}, \bibinfo{pages}{1977–1986}.
\newblock
\showISBNx{9781450390965}


\bibitem[\protect\citeauthoryear{Wei, Wang, and Zhu}{Wei et~al\mbox{.}}{2021}]%
        {AutoIAS}
\bibfield{author}{\bibinfo{person}{Zhikun Wei}, \bibinfo{person}{Xin Wang},
  {and} \bibinfo{person}{Wenwu Zhu}.} \bibinfo{year}{2021}\natexlab{}.
\newblock \showarticletitle{AutoIAS: Automatic Integrated Architecture Searcher
  for Click-Trough Rate Prediction}. In \bibinfo{booktitle}{\emph{{CIKM} '21:
  The 30th {ACM} International Conference on Information and Knowledge
  Management}}. \bibinfo{publisher}{{ACM}}, \bibinfo{address}{Australia},
  \bibinfo{pages}{2101--2110}.
\newblock


\bibitem[\protect\citeauthoryear{Weinberger, Dasgupta, Langford, Smola, and
  Attenberg}{Weinberger et~al\mbox{.}}{2009}]%
        {featurehashing}
\bibfield{author}{\bibinfo{person}{Kilian Weinberger}, \bibinfo{person}{Anirban
  Dasgupta}, \bibinfo{person}{John Langford}, \bibinfo{person}{Alex Smola},
  {and} \bibinfo{person}{Josh Attenberg}.} \bibinfo{year}{2009}\natexlab{}.
\newblock \showarticletitle{Feature Hashing for Large Scale Multitask
  Learning}. In \bibinfo{booktitle}{\emph{Proceedings of the 26th Annual
  International Conference on Machine Learning}} (Montreal, Quebec, Canada)
  \emph{(\bibinfo{series}{ICML '09})}. \bibinfo{publisher}{Association for
  Computing Machinery}, \bibinfo{address}{New York, NY, USA},
  \bibinfo{pages}{1113–1120}.
\newblock
\showISBNx{9781605585161}
\urldef\tempurl%
\url{https://doi.org/10.1145/1553374.1553516}
\showDOI{\tempurl}


\bibitem[\protect\citeauthoryear{Yan, Wang, Liu, Lin, Lee, Xu, and Zheng}{Yan
  et~al\mbox{.}}{2021a}]%
        {binarycode}
\bibfield{author}{\bibinfo{person}{Bencheng Yan}, \bibinfo{person}{Pengjie
  Wang}, \bibinfo{person}{Jinquan Liu}, \bibinfo{person}{Wei Lin},
  \bibinfo{person}{Kuang-Chih Lee}, \bibinfo{person}{Jian Xu}, {and}
  \bibinfo{person}{Bo Zheng}.} \bibinfo{year}{2021}\natexlab{a}.
\newblock \bibinfo{booktitle}{\emph{Binary Code Based Hash Embedding for
  Web-Scale Applications}}.
\newblock \bibinfo{publisher}{Association for Computing Machinery},
  \bibinfo{address}{New York, NY, USA}, \bibinfo{pages}{3563–3567}.
\newblock
\showISBNx{9781450384469}


\bibitem[\protect\citeauthoryear{Yan, Wang, Zhang, Lin, Lee, Xu, and Zheng}{Yan
  et~al\mbox{.}}{2021b}]%
        {AMTL}
\bibfield{author}{\bibinfo{person}{Bencheng Yan}, \bibinfo{person}{Pengjie
  Wang}, \bibinfo{person}{Kai Zhang}, \bibinfo{person}{Wei Lin},
  \bibinfo{person}{Kuang{-}Chih Lee}, \bibinfo{person}{Jian Xu}, {and}
  \bibinfo{person}{Bo Zheng}.} \bibinfo{year}{2021}\natexlab{b}.
\newblock \showarticletitle{Learning Effective and Efficient Embedding via an
  Adaptively-Masked Twins-based Layer}. In \bibinfo{booktitle}{\emph{{CIKM}
  '21: The 30th {ACM} International Conference on Information and Knowledge
  Management}}. \bibinfo{publisher}{{ACM}}, \bibinfo{address}{Australia},
  \bibinfo{pages}{3568--3572}.
\newblock


\bibitem[\protect\citeauthoryear{Yuan, Savarese, and Maire}{Yuan
  et~al\mbox{.}}{2021}]%
        {Cont_Spar}
\bibfield{author}{\bibinfo{person}{Xin Yuan}, \bibinfo{person}{Pedro
  Henrique~Pamplona Savarese}, {and} \bibinfo{person}{Michael Maire}.}
  \bibinfo{year}{2021}\natexlab{}.
\newblock \showarticletitle{Growing Efficient Deep Networks by Structured
  Continuous Sparsification}. In \bibinfo{booktitle}{\emph{9th International
  Conference on Learning Representations, {ICLR} 2021}}.
  \bibinfo{publisher}{OpenReview.net}, \bibinfo{address}{Austria}.
\newblock
\urldef\tempurl%
\url{https://openreview.net/forum?id=wb3wxCObbRT}
\showURL{%
\tempurl}


\bibitem[\protect\citeauthoryear{Zhang, Liu, Xie, Ktena, Tejani, Gupta, Myana,
  Dilipkumar, Paul, Ihara, Upadhyaya, Huszar, and Shi}{Zhang
  et~al\mbox{.}}{2020}]%
        {doublehashing}
\bibfield{author}{\bibinfo{person}{Caojin Zhang}, \bibinfo{person}{Yicun Liu},
  \bibinfo{person}{Yuanpu Xie}, \bibinfo{person}{Sofia~Ira Ktena},
  \bibinfo{person}{Alykhan Tejani}, \bibinfo{person}{Akshay Gupta},
  \bibinfo{person}{Pranay~Kumar Myana}, \bibinfo{person}{Deepak Dilipkumar},
  \bibinfo{person}{Suvadip Paul}, \bibinfo{person}{Ikuhiro Ihara},
  \bibinfo{person}{Prasang Upadhyaya}, \bibinfo{person}{Ferenc Huszar}, {and}
  \bibinfo{person}{Wenzhe Shi}.} \bibinfo{year}{2020}\natexlab{}.
\newblock \bibinfo{booktitle}{\emph{Model Size Reduction Using Frequency Based
  Double Hashing for Recommender Systems}}.
\newblock \bibinfo{publisher}{Association for Computing Machinery},
  \bibinfo{address}{New York, NY, USA}, \bibinfo{pages}{521–526}.
\newblock
\showISBNx{9781450375832}


\bibitem[\protect\citeauthoryear{Zhang, Du, and Wang}{Zhang
  et~al\mbox{.}}{2016}]%
        {FNN}
\bibfield{author}{\bibinfo{person}{Weinan Zhang}, \bibinfo{person}{Tianming
  Du}, {and} \bibinfo{person}{Jun Wang}.} \bibinfo{year}{2016}\natexlab{}.
\newblock \showarticletitle{Deep Learning over Multi-field Categorical Data - -
  {A} Case Study on User Response Prediction}. In
  \bibinfo{booktitle}{\emph{Advances in Information Retrieval - 38th European
  Conference on {IR} Research, {ECIR} 2016}}, Vol.~\bibinfo{volume}{9626}.
  \bibinfo{publisher}{Springer}, \bibinfo{address}{Italy},
  \bibinfo{pages}{45--57}.
\newblock
\urldef\tempurl%
\url{https://doi.org/10.1007/978-3-319-30671-1\_4}
\showDOI{\tempurl}


\bibitem[\protect\citeauthoryear{Zhao, Liu, Liu, Tang, Guo, Shi, Wang, Gao, and
  Long}{Zhao et~al\mbox{.}}{2021}]%
        {AutoDim}
\bibfield{author}{\bibinfo{person}{Xiangyu Zhao}, \bibinfo{person}{Haochen
  Liu}, \bibinfo{person}{Hui Liu}, \bibinfo{person}{Jiliang Tang},
  \bibinfo{person}{Weiwei Guo}, \bibinfo{person}{Jun Shi},
  \bibinfo{person}{Sida Wang}, \bibinfo{person}{Huiji Gao}, {and}
  \bibinfo{person}{Bo Long}.} \bibinfo{year}{2021}\natexlab{}.
\newblock \showarticletitle{AutoDim: Field-aware Embedding Dimension Searchin
  Recommender Systems}. In \bibinfo{booktitle}{\emph{{WWW} '21: The Web
  Conference 2021}}. \bibinfo{publisher}{{ACM} / {IW3C2}},
  \bibinfo{address}{Slovenia}, \bibinfo{pages}{3015--3022}.
\newblock


\bibitem[\protect\citeauthoryear{Zhu, Liu, Yang, Zhang, and He}{Zhu
  et~al\mbox{.}}{2021}]%
        {fuxictr}
\bibfield{author}{\bibinfo{person}{Jieming Zhu}, \bibinfo{person}{Jinyang Liu},
  \bibinfo{person}{Shuai Yang}, \bibinfo{person}{Qi Zhang}, {and}
  \bibinfo{person}{Xiuqiang He}.} \bibinfo{year}{2021}\natexlab{}.
\newblock \showarticletitle{Open Benchmarking for Click-Through Rate
  Prediction}. In \bibinfo{booktitle}{\emph{Proceedings of the 30th ACM
  International Conference on Information \& Knowledge Management}}.
  \bibinfo{publisher}{Association for Computing Machinery},
  \bibinfo{address}{Australia}, \bibinfo{pages}{2759–2769}.
\newblock


\end{thebibliography}


\end{document}